\documentclass[pra,twocolumn,longbibliography,superscriptaddress,10pt]{revtex4-1}
\usepackage{graphicx}
\usepackage{dcolumn}
\usepackage{color}
\usepackage{times}
\usepackage{bm}
\usepackage{amssymb}
\usepackage{amsmath}
\usepackage{epsfig}
\usepackage{epstopdf}
\usepackage{dsfont}
\usepackage{subfigure}
\usepackage{tikz}
\usepackage[colorlinks, citecolor=blue, linkcolor=blue, urlcolor=blue]{hyperref}
\usepackage[mathscr]{euscript}
\usepackage{multirow}
\usepackage{tabularx}
\usepackage{tikz}
\usetikzlibrary{decorations.pathreplacing, calc}
\makeatletter
\renewcommand{\maketag@@@}[1]{\hbox{\m@th\normalsize\normalfont#1}}
\makeatother

\begin{document}
\renewcommand{\thefootnote}{\fnsymbol {footnote}}

\title{Generalized entropic uncertainty relation and non-classicality in Schwarzschild black hole}

\author{Rui-Jie Yao}
\affiliation{School of Physics \& Optoelectronic Engineering, Anhui University, Hefei 230601,  People's Republic of China}

\author{Dong Wang} \email{dwang@ahu.edu.cn}
\affiliation{School of Physics \& Optoelectronic Engineering, Anhui University, Hefei 230601,  People's Republic of China}

\date{\today}

\begin{abstract}
{The uncertainty principle constitutes a fundamental pillar of quantum theory, representing one of the most distinctive features that differentiates quantum mechanics from classical physics. In this study, we firstly propose a novel generalized entropic uncertainty relation (EUR) for arbitrary multi-measurement in the many-body systems, and rigorously derive a significantly tighter bound compared to existing formulations. Specifically, we discuss the proposed EUR in the context of Schwarzschild black hole, where we demonstrate the superior tightness of our derived bound. The study further elucidates the dynamical evolution of multipartite quantum coherence and entanglement in the curved spacetime. A particularly noteworthy finding reveals the exact equivalence between entanglement and $l_1$-norm coherence for arbitrary $N$-partite Greenberger-Horne-Zeilinger-type (GHZ-type) states. Moreover, we find that quantum coherence is significantly diminished and the measurement uncertainty increases to a stable maximum with increasing Hawking temperature. Thus, the findings of this study contribute to a deeper understanding of non-classicality and quantum resources in black holes.}\end{abstract}

\maketitle
\section{Introduction}
\label{sec1}
General relativity first predicted the formation of black holes, which are associated with the gravitational collapse of sufficiently massive stars \cite{SWH}. In fact, there are varieties of curved spacetime, including the Schwarzschild spacetime, the Kerr-Newman spacetime, and the Garfinkle-Horowitz-Strominger spacetime \cite{ETN,DG}. It is well known that the Schwarzschild black hole is the simplest model of the black hole, describing a black hole with no electric charge and no rotation. The Schwarzschild black hole is the first static solution based on Einstein's equations of general relativity. In order to better understand the nature of the curved spacetime, it is required to figure out the properties of its quantumness. 

The uncertainty principle is one of the most remarkable features in quantum information theory, which demonstrates the intrinsic uncertainty of nature from an information-theory perspective. It was firstly proposed by Heisenberg \cite{WH} and was certified by Kennard \cite{EHK} with respect to position {$\hat x$} and momentum {$\hat p$}
\begin{equation}
\begin{split}
 \Delta \hat x \Delta \hat p \ge \frac{\hbar }{2},
\label{Eq.1.1}
\end{split}
\end{equation}
where {$\Delta \hat x = \sqrt {\left\langle {{\hat x^2}} \right\rangle  - {{\left\langle \hat x \right\rangle }^2}} $} represents the standard deviation with ${\left\langle \hat x \right\rangle }$ representing the average of $\hat x$. Subsequently, Robertson generalized it to the form of two arbitrary noncommutative observables.

In 1957, Everett \cite{HE} and Hirschman \cite{IIH} brought entropy into the  uncertainty principle, and then Deutsch \cite{DD} proposed a generalization of the entropic uncertainty relation (EUR) as
\begin{equation}
\begin{split}
H(\hat{R})+H(\hat{Q})\ge \log_2 \left ( \frac{2}{1+\sqrt{c\left ( \hat{R},\hat{Q} \right ) } }  \right ) ^2,
\label{Eq.1.2}
\end{split}
\end{equation}
where {$H(\hat{X})=-\sum_i p_i \log _2 p_i$} stands for Shannon entropy with {$p_i$} being the probability of the {$i$}-th outcome. The maximal overlap {$c\left ( \hat{R},\hat{Q} \right )= \underset{i,j}{\max} \left | \left \langle {\psi_i^{\hat{R}} } \middle| {\psi_j^{\hat{Q}}}  \right \rangle  \right |^2 $} where {$|{\psi_i^{\hat{R}} }\rangle$ ($ | {\psi_j^{\hat{Q}}} \rangle) $} is the eigenvector of observable {$\hat{R} \ (\hat{Q})$}. Later, based on Kraus' conjecture \cite{KK}, Maassen and Uffink \cite{HM} optimized the relation as
\begin{equation}
\begin{split}
H(\hat{R})+H(\hat{Q})\ge - \ln c \left ( \hat{R},\hat{Q} \right ).
\label{Eq.1.3}
\end{split}
\end{equation}
Additionally, it was interestingly found the uncertainty can be reduced when the measured particle is correlated to another particle. Recently, Renes  {\it et al.} \cite{JMR} and Berta  $et$ $al.$ \cite{MB} proposed quantum-memory-assisted entropic uncertainty relations (QMA-EURs), which can be expressed as
\begin{equation}
\begin{split}
S( \hat{R} | B ) + S( \hat{Q} | B ) \geqslant q_{MU} + S(A|B) \quad  (\mathrm{for\  biparticle\ {\hat\rho _{AB}}}),
\label{Eq.1.4}
\end{split}
\end{equation}\\
and
\begin{equation}
\begin{split}
S( \hat{R} | B ) + S( \hat{Q} | C ) \geqslant q_{MU}\quad  (\mathrm{for\  triparticle\  {\hat\rho _{ABC}}}).
\label{Eq.1.5}
\end{split}
\end{equation}
Where $B$ and $C$ represent the quantum memories, and {$S( \hat{R} | B ) = S(\hat\rho_{\hat{R}B}) - S(\hat\rho_{B})$} denotes the conditional Von Neumann entropic of the system after measuring {$\hat{R}$} on {$A$} and the post-measurement states can be expressed as: {${{\hat\rho}_{\hat{R}B}} = \sum_{i} {\left( {{{\left| {\psi _i^{\hat{R}}} \right\rangle }}\left\langle {\psi _i^{\hat{R}}} \right| \otimes {\mathds{I}_B}} \right){{\rho}_{AB}}\left( {{{\left| {\psi _i^{\hat{R}}} \right\rangle }}\left\langle {\psi _i^{\hat{R}}} \right| \otimes {\mathds{I}_B}} \right)}$}, and $q_{MU} = - \log_2 c \left ( \hat{R},\hat{Q} \right )$ is denoted \cite{JMR,MB}. In recent years, many authors have made great effort to present tighter EURs \cite{MAB,SW,AKP,MLH,MNB,TP,LM,PJC,LR1,SZ,LR2,JZ,SL,YLX,FA,DW,FM,BFX,LJL,LW,QHZ,TYW,SZLJL,ZAW}. For examples, Adabi {\it et al.} \cite{FA} optimized the EUR by considering Holevo quantites and mutual information, and Wu {\it et al.} \cite{LW} proposed tighter generalized entropic uncertainty relations in multipartite systems. With the development of quantum information technologies, EURs have been widely applied to various quantum information processing tasks, including quantum metrology \cite{CSY,VG}, quantum teleportation \cite{ML}, quantum key distribution \cite{NJC}, quantum steering \cite{HMW,KS,JS}, and so on \cite{JMR2,JMR3,GJ,DWF}.
Besides, EURs for multiple memories and measurements have attracted much attention.
 Ming {\it et al.} \cite{FM} proposed an improved tripartite QMA-EUR by incorporating mutual information and Holevo quantity in the presence of quantum memories
\begin{equation}
\begin{split} 
S\left( {{{\hat M}_1}|{B}} \right) + S\left( {{{\hat M}_2}|{C}} \right) \ge  - {\log _2}{C_{ij}} +   \max\{ 0,{\delta _1}\}, 
\label{Eq.1.6}
\end{split}
\end{equation}
where $\delta_{1} = 2S(A) + q_{MU} - \mathcal{I}(A:B) - \mathcal{I}(A:C) + \mathcal{I}(\hat M_{2}:B) + \mathcal{I}(\hat M_{1}:C) - H(\hat M_{1}) - H(\hat M_{2})$. Later, Wu {\it et al.} \cite{LW} improved the above bound
further
\begin{equation}
\begin{split} 
\sum_{i=1}^m S(\hat{M}_i | B_i) \geq -\frac{1}{m-1} \log_2 \left( \prod_{\substack{i \neq j}}^m c_{ij} \right) + \max\{0, \delta_2\},
\label{Eq.1.7}
\end{split}
\end{equation}
with 
\begin{equation}
\begin{split} 
\delta_2& = -\frac{1}{m-1} \log_2 \left( \prod_{\substack{i \neq j}}^m c_{ij} \right) + nS(A)\\
& - \sum_{i=1}^m H(\hat{M}_i) - \sum_{i=1}^m I(\hat{M}_i : B_i)
\nonumber
\end{split}
\end{equation}
where ${M_i}$ denotes the $i$-th measurement on subsystem $A$ and ${B_i}$ represents the $i$-th quantum memory in the multipartite system.

Herein, we rigorously derive a generalized entropic uncertainty relation for arbitrary multiple observables in many-body systems through the framework of Holevo quantity. This theoretical advancement establishes a novel uncertainty bound that partially extends conventional formulations. To demonstrate its practical utility, we implement this generalized uncertainty principle in the context of Schwarzschild black hole, thereby elucidating the dynamical evolution of quantum uncertainties in the curved spacetime. Furthermore, we systematically investigate the intrinsic quantum features of spacetime geometry through the lens of quantum correlations and coherence measures. Notably, our analysis reveals a remarkable equivalence between spacetime entanglement and quantum coherence for $N$-particle GHZ-type states - a finding that bridges quantum information theory with relativistic gravity frameworks. This work establishes a critical connection between quantum measurement theory, many-body physics, and gravitational physics through the prism of information-theoretic principles.

The paper is organized as follows: in Section \ref{sec2}, we firstly present a rigorous derivation of a generalized entropic uncertainty relation (EUR) applicable to multi-observable scenarios in many-body quantum systems, establishing a unified theoretical framework that transcends existing formulations. In Section \ref{sec3}, we introduce the vacuum structure of Dirac particles in Schwarzschild spacetime. In Section \ref{sec4}, we conduct detailed examinations of two fundamental quantum resources - entropic uncertainty dynamics and $N$-partite quantum concurrence - in the context of evaporating Schwarzschild black holes, establishing their intrinsic connection through relativistic quantum information theory.  Finally, our paper ends with a concise conclusion.

\section{Generalized entropic uncertainty relations for multiple measurements within multiparty}
\label{sec2}
In this section, we will focus on presenting a generalized entropic uncertainty relation and providing the detailed proof.
\subsection{Optimized tripartite entropic uncertainty relation }
\label{sec2.1}
Firstly, we rewrite the QMA-EURs in eqs. (\ref{Eq.1.4}) and (\ref{Eq.1.5}) as
\begin{equation}
S\left( {{\hat M}_i}|{B_t} \right) + S\left( {{\hat M}_j}|{B_s} \right) \ge  - {\log _2}{C_{ij}} + S\left( {A|{B_t}} \right) \quad (t = s),
\label{Eq.201}
\end{equation}
and
\begin{equation}
S\left( {{\hat M}_i}|{B_t} \right) + S\left( {{\hat M}_j}|{B_s} \right) \ge  - {\log _2}{C_{ij}} \quad (t \ne s).
\label{Eq.202}
\end{equation}
Where $C_{ij}$ is the maximal overlap of $\hat{M}_i$ and $\hat{M}_j$. In the absence of quantum storage $B$, we have \cite{MB}
\begin{equation}
S\left( {{{\hat M}_i}} \right) + S\left( {{{\hat M}_j}} \right) \ge  - {\log _2}{C_{ij}} + S\left( A \right).
\label{Eq.203}
\end{equation}
Meanwhile, the conditional entropy ${S(\hat M|B)}$ and Holevo quantity ${H(\hat M:B)}$ satisfy the following equality
\begin{equation}
S(\hat M) = S(\hat M|B) + H(\hat M:B).
\label{Eq.204}
\end{equation}
As a result, the inequality Eq. (\ref{Eq.202}) can be improved as
\begin{equation}
\begin{split}
S(&\mathop {{{\hat M}_i}|{B_t}}\limits^{} ) + S(\mathop {{{\hat M}_j}|{B_s}}\limits^{} )\\
&= S({{\hat M}_i}) + S({{\hat M}_j}) - H({{\hat M}_i}:{B_t}) - H({{\hat M}_j}:{B_s})\\
 &\ge - {\log _2}{C_{ij}} + S\left( A \right) - H({{\hat M}_i}:{B_t}) - H({{\hat M}_j}:{B_s}).
\label{Eq.205}
\end{split}
\end{equation}

\subsection{Universal EUR for multiple measurements in multipartite system}
\label{sec2.2}
Suppose there are $n+1$ legitimate participators: Alice, Bob$_1$, Bob$_2$,..., Bob$_n$, who share a $n+1$ - partite quantum state.  Alice randomly selects a measurement from the set of measurements ${\cal M} = \left\{ {{\hat M_i}} \right\}(i = 1,2,...,m)$. ${\cal M}$ has $n$ nonempty subsets and satisfies the following relation: $\bigcup\limits_{t = 1}^n {{S_t} = {\cal M}}$ and ${S_s} \cap {S_t} = \emptyset$ \: for $s \ne t$. For subset ${S_t}$, the dimension is ${m_t}$ $(t=1,2,...,n)$ and satisfies the relation: $\sum\limits_{t = 1}^n {{m_t} = m}$. In other words, ${S_1} = \{ {\hat M_i}\},(i = 1,2,...,{m_1})$, ${S_2} = \{ {\hat M_i}\} ,(i = {m_1} + 1,{m_1} + 2,...,{m_1} + {m_2})$, ..., ${S_n} = \{ {\hat M_i}\} ,(i = {m_1} + \cdots + {m_{n - 1}} + 1,{m_1} + \cdots + {m_{n - 1}} + 2,...,{m_1} + \cdots + {m_n})$. According to the principle of linear addition of inequalities, we present the following simply constructed bound (SCB) for quantum-memory-assisted entropic uncertainty relations

\emph{Lemma 1}: We firstly present the generalized EUR without Holevo quantity for $m$ measurements and $n$ memories as
\begin{equation}
\begin{split}
\sum_{t = 1}^n \sum_{{M_i} \in {S_t}} S(\hat{M_i}|B_t) \ge &- \frac{\sum\limits_{i = 1}^{m - 1} \sum\limits_{j = i + 1}^{m} \log_2 c_{i,j}}{m - 1}\\
& + \frac{\sum\limits_{i = 1}^n \left[ \frac{m_i(m_i - 1)}{2} \right] S(A|B_i)}{m - 1}.
\label{Eq.206}
\end{split}
\end{equation}

\emph{Proof}: We can sum   Eqs. (\ref{Eq.201}) and (\ref{Eq.202}) for all $i < j$ and normalize on both sides. Thus,  the Lemma 1 is proofed.

\emph{Theorem}: By considering the Holevo quantity, the generalized EUR for arbitrary multi-observable in multipartite system can be expressed as
\begin{equation}
\begin{split}
\sum_{t = 1}^n \sum_{{M_i} \in {S_t}}& S(\hat{M_i}|B_t) \ge - \frac{\sum\limits_{i = 1}^{m - 1} \sum\limits_{j = i + 1}^{m} \log_2 c_{i,j}}{m - 1} \\ 
\quad &+ \frac{\sum\limits_{i = 1}^n \left[ \frac{m_i(m_i - 1)}{2} \right] S(A|B_i)}{m - 1} + \max\{0,\delta\},
\label{Eq.207}
\end{split}
\end{equation}
where
\begin{equation}
\begin{split}
\delta =  &- \frac{\sum\limits_{t = 1}^n \sum\limits_{{M_i} \in {S_t}} \left[ \left( \sum\limits_{j = 1,j \ne t}^n m_j \right)H(\hat{M_i}:B_t) \right]}{m - 1}\\
&+ \frac{\sum\limits_{i = 1}^{n - 1} m_i \left( \sum\limits_{j = i + 1}^n m_j \right)S(A)}{m - 1}.
\nonumber
\end{split}
\end{equation}

\emph{Proof}: Firstly, by considering  the measuring subset $S_1$ and quantum memory $B_1$, we can write all the $({m_1 - 1})$ inequalities for two measurements ${\hat M_i}$ and ${\hat M_j}$ $(i \ne j)$ as
\begin{equation}
\begin{array}{l}
\begin{split}
S(\mathop {{{\hat M}_1}|{B_1}}\limits^{} ) + S(\mathop {{{\hat M}_2}|{B_1}}\limits^{} ) &\ge  - {\log _2}{C_{1,2}} + S\left( {A|{B_1}} \right),\\
&\vdots \\
S(\mathop {{{\hat M}_1}|{B_1}}\limits^{} ) + S(\mathop {{{\hat M}_{{m_1}}}|{B_1}}\limits^{} ) &\ge  - {\log _2}{C_{1,{m_1}}} + S\left( {A|{B_1}} \right),
\label{Eq.208}
\end{split}
\end{array}
\end{equation}
in terms of the EUR in Eq. (\ref{Eq.201}). Then, we consider the situations of different quantum memories $B_1$ and $B_s$ ${(s > 1)}$. Similar to the above inequalities, we apply the corresponding measurement subset $S_i$ on quantum memory $B_i$ ($i = 1, 2, ..., n$). Resorting to Eq.  (\ref{Eq.205}), we obtain 
\begin{equation}
\begin{array}{l}
\begin{split}
S(\mathop {{{\hat M}_1}|{B_1}}\limits^{} ) + S(\mathop {{{\hat M}_{{m_1} + 1}}|{B_2}}\limits^{} ) \ge&  - {\log _2}{C_{1,{m_1} + 1}} + S\left( A \right)\\
& - H({{\hat M}_1}:{B_1})\\
& - H({{\hat M}_{{m_1} + 1}}:{B_2}),\\
 &\vdots \\
S(\mathop {{{\hat M}_1}|{B_1}}\limits^{} ) + S(\mathop {{{\hat M}_{{m_1} + {m_2}}}|{B_2}}\limits^{} ) \ge&  - {\log _2}{C_{1,{m_1} + {m_2}}} + S\left( A \right)\\
& - H({{\hat M}_1}:{B_1})\\
& - H({{\hat M}_{{m_1} + {m_2}}}:{B_2}),\\
 &\vdots \\
S(\mathop {{{\hat M}_1}|{B_1}}\limits^{} ) + S(\mathop {{{\hat M}_{{m_1} + \cdots + {m_n}}}|{B_n}}\limits^{} ) \ge&  - {\log _2}{C_{1,{m_1} + \cdots + {m_n}}}\\
& + S\left( A \right) - H({{\hat M}_1}:{B_1}) \\
& - H({{\hat M}_{{m_1} + \cdots + {m_n}}}:{B_n}).
\label{Eq.209}
\end{split}
\end{array}
\end{equation}
By adding the above inequalities and dividing both sides by ${({m_1} +  \cdots  + {m_n} - 1)}$, we can obtain
\begin{equation}
\begin{split}
S({\hat M_1}|&{B_1}) + \frac{{\sum\limits_{i = 2}^{{m_1}} {S({{\hat M}_i}|{B_1})}  + \sum\limits_{t = 2}^n {\sum\limits_{{M_i} \in {S_t}} S } ({{\hat M}_i}|{B_t}){\rm{ }}}}{{({m_1} +  \cdots  + {m_n} - 1)}}\\
&\ge \frac{{ - \sum\limits_{j = 2}^{{m_1} +  \cdots  + {m_n}} {{{\log }_2}{C_{1,j}}}  + ({m_1} - 1)S(A|{B_1})}}{{({m_1} +  \cdots  + {m_n} - 1)}}\\
&\quad + \frac{{ ({m_2} +  \cdots  + {m_n})S(A)}}{{({m_1} +  \cdots  + {m_n} - 1)}}\\
&\quad - \frac{{({m_2} +  \cdots  + {m_n})H({{\hat M}_1}:{B_1})}}{{({m_1} +  \cdots  + {m_n} - 1)}} \\
&\quad  -\frac{{\sum\limits_{t = 2}^n {\sum\limits_{{M_i} \in {S_t}} {H({{\hat M}_i}:{B_t})} } }}{{({m_1} +  \cdots  + {m_n} - 1)}} . 
\label{Eq.210}
\end{split}
\end{equation}
Likewise, one can easily attain the following results for $i = 2,..., m_1$
\begin{equation}
\begin{split}
S({\hat M_2}&|{B_1}) + \frac{{\sum\limits_{i = 3}^{m_1} {S({{\hat M}_i}|{B_1})}  + \sum\limits_{t = 2}^n {\sum\limits_{{M_i} \in {S_t}} S } ({{\hat M}_i}|{B_t})}}{{({m_1} +  \cdots  + {m_n} - 2)}}\\
&\ge \frac{-{\sum\limits_{j = 3}^{{m_1} +  \cdots  + {m_n}} {{{\log }_2}{C_{2,j}}} }+{({m_1} - 2)S(A|{B_1})}}{{({m_1} +  \cdots  + {m_n} - 2)}}\\
&\quad + \frac{({m_2} +  \cdots  + {m_n})S(A)}{{({m_1} +  \cdots  + {m_n} - 2)} }\\
&\quad - \frac{{({m_2} +  \cdots  + {m_n})H({{\hat M}_2}:{B_1}) }}{{({m_1} +  \cdots  + {m_n} - 2)}}\\
&\quad -\frac{{\sum\limits_{t = 2}^n {\sum\limits_{{M_i} \in {S_t}} {H({{\hat M}_i}:{B_t})} } }}{{({m_1} +  \cdots  + {m_n} - 2)}}.
\label{Eq.211}
\end{split}
\end{equation}
\centerline{\vdots} 
\begin{equation}
\begin{split}
S({\hat M_{{m_1}}}|{B_1}) &+ \frac{{\sum\limits_{t = 2}^n {\sum\limits_{{M_i} \in {S_t}} S } ({{\hat M}_i}|{B_t})}}{{({m_2} +  \cdots  + {m_n})}}{\rm{ }}\\
&\ge - \frac{{\sum\limits_{j = {m_1} + 1}^{{m_1} +  \cdots  + {m_n}} {{{\log }_2}{C_{{m_1},j}}} }}{{({m_2} +  \cdots  + {m_n})}} + S(A)\\
&\quad -\frac{{({m_2} +  \cdots  + {m_n})H({{\hat M}_{{m_1}}}:{B_1})}}{{({m_2} +  \cdots  + {m_n})}}\\
&\quad -\frac{{\sum\limits_{t = 2}^n {\sum\limits_{{M_i} \in {S_t}} {H({{\hat M}_i}:{B_t})} } }}{{({m_2} +  \cdots  + {m_n})}}. 
\label{Eq.212}
\end{split}
\end{equation}
Summarizing these inequalities in the case of $B_t = B_1$ and ${\hat M_i} \in {S_1}$, we can get 
\begin{equation}
\begin{split}
\sum\limits_{k = 1}^{m_1} &S(\hat{M_k}|B_1) + \frac{m_1 \sum\limits_{t = 2}^n \sum\limits_{M_i \in S_t} S(\hat{M_i}|B_t)}{m_1 + \cdots + m_n - 1} \\
&\ge -\frac{\sum\limits_{i = 1}^{m_1} \sum\limits_{j = i + 1}^{m_1 + \cdots + m_n} \log_2 c_{i,j}}{m_1 + \cdots + m_n - 1} \\
&\quad + \frac{m_1(m_2 + \cdots + m_n) S(A) + \frac{m_1(m_1 - 1)}{2} S(A|B_1)}{m_1 + \cdots + m_n - 1} \\
&\quad - \frac{(m_2 + \cdots + m_n) \sum\limits_{i = 1}^{m_1} H(\hat{M_i}:B_1)}{m_1 + \cdots + m_n - 1} \\
&\quad - \frac{m_1 \sum\limits_{t = 2}^n \sum\limits_{M_i \in S_t} H(\hat{M_i}:B_t)}{m_1 + \cdots + m_n - 1}.
\label{Eq.213}
\end{split}
\end{equation}

Similar to Eq. (\ref{Eq.213}), we consider the case where quantum storage is ${B_2}$ and measurement is subset ${S_2}$, that is, $B_t = B_2$ and ${\hat M_i} \in {S_2}$. The following inequality can be obtained as 
\begin{equation}
\begin{split}
\sum\limits_{k = {m_1} + 1}^{{m_1} + {m_2}} &S ({\hat M_k}|{B_2}) + \frac{{{m_2}\sum\limits_{t = 2}^n {\sum\limits_{{M_i} \in {S_t}} S } (\hat {{M_i}}|{B_t})}}{{({m_2} + ... + {m_n} - 1)}}\\
&\ge  - \frac{{\sum\limits_{i = {m_1} + 1}^{{m_1} + {m_2}} {\sum\limits_{j = i + 1}^{{m_1} +  \cdots  + {m_n}} {{{\log }_2}} } {c_{i,j}}}}{{({m_2} + ... + {m_n} - 1)}}\\
&\quad + \frac{{m_2}({m_3} + ... + {m_n})S(A) + {\frac{{{m_2}({m_2} - 1)}}{2}S(A|{B_2})}}{{({m_2} + ... + {m_n} - 1)}} \\
&\quad - \frac{({m_3} + ... + {m_n})\sum\limits_{i = {m_1} + 1}^{{m_1} + {m_2}} {H({{\hat M}_i}:{B_3})}}{{({m_2} + ... + {m_n} - 1)}}\\
&\quad - \frac{{m_2}\sum\limits_{t = 3}^n {\sum\limits_{{M_i} \in {S_t}} {H({{\hat M}_i}:{B_t})}}}{{({m_2} + ... + {m_n} - 1)}}. 
\label{Eq.214}
\end{split}
\end{equation}
In this way, one can work out the inequalities for $B_t = B_3,...,B_{n-1}$ as shown above. Specifically, the inequality for $B_t=B_n$ is as follows
\begin{equation}
\begin{split}
\sum\limits_{k = {m_1} +  \cdots  + {m_{n - 1}} + 1}^{{m_1} +  \cdots  + {m_n}}& S ({{\hat M}_k}|{B_n})\ge \frac{{m_n}}{2}S(A|{B_n})\\
&  - \frac{\sum\limits_{i = {m_1} + \cdots + {m_{n - 1}} + 1}^{{m_1} + \cdots + {m_n} - 1} {\sum\limits_{j = i + 1}^{{m_1} +  \cdots  + {m_n}} {{{\log }_2}} } {c_{i,j}}}{{m_n} - 1}.
\label{Eq.215}
\end{split}
\end{equation}

Finally, by summing the inequality from $B_t=B_1$ to $B_t=B_n$ and subsequently dividing both sides by the normalization factor $(m_1+\cdots+m_n - 1)$, we rigorously derive Eq. (\ref{Eq.207}), thereby completing the proof. In order to elucidate the  process of proof in Eq. (\ref{Eq.207}), Appendix \ref{appendixA} has been provided for some details.
It  emphasizes that the maximal overlap $C_{ij}$ between any two measurements $\hat M_i$ and $\hat M_j$, defined as $C_{ij} = \max_{k,l} |\langle \psi_k^{\hat M_i} | \psi_l^{\hat M_j} \rangle|^2$, is a property of the measurements themselves. This definition holds independently of whether $\hat M_i$ and $\hat M_j$ belong to the same subset $S_t$ or to different subsets in the above partition.

We present a comparison of the performance between our QMA-EUR and the previous ones \cite{QHZ}. For the case of tripartite system ${\hat \rho _{ABC}}$ with three measurements ${\{ {\hat M_1},{\hat M_2},{\hat M_3}\} }$. Alice (holding particle $A$) performs three measurements on $A$ and two quantum memories are held by Bob and Charlie, with measurement subsets partitioned as ${S_1 = \{\hat M_1\} }$ and ${S_2 = \{\hat M_2, \hat M_3\}}$. In this case, our EUR reduces to
\begin{equation}
\begin{split}  
S({\hat M_1}&|B) + S({\hat M_2}|C) + S({\hat M_3}|C)\\
& \ge -\frac{1}{2}{\log _2}(\mathop \prod \limits_{i < j}^3 {c_{i,j}})  + \frac{1}{2}S(A|C) + \max \left\{ {0,{\delta _1}} \right\},
\label{Eq.216}
\end{split}
\end{equation}
where ${\delta _1 = S(A) - H({{\hat M}_1}:B) - \frac{1}{2}[H({{\hat M}_2}:C) + H({{\hat M}_3}:C)]}$. To illustrate the performance of our EUR, we rewrite the result in Ref. \cite{QHZ} as   
\begin{equation}
\begin{split}  
S({\hat M_1}&|B) + S({\hat M_2}|C) + S({\hat M_3}|C)\\
& \ge -\frac{1}{2}{\log _2}(\mathop \prod \limits_{i < j}^3 {c_{i,j}}) + \frac{1}{2}S(A|C) + \max \left\{ {0,{\delta _2}} \right\},
\label{Eq.217}
\end{split}
\end{equation}
where ${\delta _2 = S(A) - \frac{1}{2}I(A:C) - H({{\hat M}_1}:B) - H({{\hat M}_2}:C)}$\\
${ - H({{\hat M}_3}:C)}$, and ${I(A:B) = S({\hat \rho _A}) + S({\hat \rho _B}) - S({\hat \rho _{AB}})}$ denotes the mutual information. We set
\begin{equation}
\begin{split}
\delta _1 -\delta_2 = \frac{1}{2}I(A:C) + \frac{1}{2}H({{\hat M}_2}:C) + \frac{1}{2}H({{\hat M}_3}:C).
\label{Eq.218}
\end{split}
\end{equation}
Since both mutual information and Holevo quantity are non-negative,we have ${\delta _1 \ge \delta _2}$, indicating that our bound is tighter than the previous in the current scenario.

\section{Vacuum structure of dirac particles in schwarzschild space-time}
\label{sec3}
We first go over the vacuum states of  Schwarzschild  black hole. In general, the metric of the Schwarzschild black hole can be expressed as
\begin{equation}
\begin{split}
{ds^{2} =} & {- \left(1-\frac{2M}{r}\right) dt^{2}} + \left(1-\frac{2M}{r}\right) ^{-1}dr^{2} \\&{+r^2(d\theta ^{2}+\sin^{2}\theta d\phi ^{2} ) ,}
\label{Eq.301}
\end{split}
\end{equation}
herein {$M$} is the mass of the black hole. In this work,  the mostly-plus signature is adopted. Additionally, the Hawking temperature can be expressed as  {$T_H = \frac{\kappa}{2 \pi }$}, where surface gravity is {${\kappa} = \frac{1}{4M }$}.
In this paper, we set the gravitational constant {$G$}, the reduced Planck constant {$\hbar$}, the speed of light {$c$}, and the Boltzmann constant {${k_B}$} to unit 1, for simplicity.

The massless Dirac field in this Schwarzschild spacetime is given by
\begin{equation}
\left[ {{\gamma ^a}e_a^\mu \left( {{\partial _\mu } - {\Gamma _\mu }} \right)} \right]\psi  = 0,
\label{Eq.302}
\end{equation}
where $\gamma ^a$ stands for Dirac matrices and $\Gamma _{\mu }$ is spin connection. To specify its vacuum structure for different observers placed at different distances of the event horizon, we introduce the null Kruskal-Szekeres coordinates
\begin{equation}
{u =  - {\kappa ^{ - 1}}{e^{ - \kappa \left( {t - {r^*}} \right)}},v = {\kappa ^{ - 1}}{e^{\kappa \left( {t + {r^*}} \right)}} },
\label{Eq.303}
\end{equation}
where {${r^*} = r + 2M\ln \left| {\frac{r}{{2M}} -1} \right| $} is tortoise coordinate. In terms of the new coordinates, the Schwarzschild metric is 
\begin{equation}
{d{s^2} =  - \frac{1}{{2\kappa r}}{e^{ - 2\kappa r}}dudv + {r^2}\left( {d{\theta ^2} + {{\sin }^2}\theta d{\phi ^2}} \right)}.
\label{Eq.304}
\end{equation}

Since we can analyze the vacuum structures of quantum field in three regions. The vacuum corresponding to a free falling observer (Alice) close to the event horizon is the Hartle-Hawking vacuum {$|0\rangle_H$}, which behaves locally like the Minkowski vacuum {$|0\rangle_M$}. For an observer located at a position {$r_0$} close to the event horizon who has a proper acceleration {${a_0} = \frac{M}{{{r_0}^2\sqrt {1 - \frac{{2M}}{{{r_0}}}} }}$}, the vacuum state is called Boulware vacuum {$|0\rangle_B$}, which is analogous to the Rindler vacuum {$|0\rangle_I$}. Besides, there is another Boulware vacuum {${|0\rangle _{\overline B }}$} in region IV and analogous to {$|0\rangle_{IV}$} in the Rindler case. Considering that the states consist of different frequencies modes, vacuum satisfies {$|0\rangle_H  = { \otimes _i}{\left| {{0_{{\omega _i}}}} \right\rangle _H}$} and first excitation {$|1\rangle_H  = { \otimes _i}{\left| {{1_{{\omega _i}}}} \right\rangle _H}$}. Their analogy is as follows \cite{JLH}
\begin{equation}
\begin{split}
|0\rangle_B &\leftrightarrow |0\rangle_I \leftrightarrow |0\rangle_{\text{Bob}}, \\
|0\rangle_{\overline{B}} &\leftrightarrow |0\rangle_{IV} \leftrightarrow |0\rangle_{\overline{\text{Bob}}},\\  
|0\rangle_H &\leftrightarrow |0\rangle_M \leftrightarrow |0\rangle_{\text{Alice}}.
\label{Eq.305}
\end{split}
\end{equation}
To connect flat spacetime results to the black hole scenario, we need to recall the Dirac field quantization in Rindler spacetime. We first construct specialized Minkowski modes as linear combinations of monochromatic solutions \cite{EMM1}
\begin{equation}
\psi^M_{\omega_j,\sigma} = \sum_i D_{ij} u^M_{\hat{\omega}_i,\sigma}, \quad \bar{\psi}^M_{\omega_j,\sigma} = \sum_i E_{ij} v^M_{\hat{\omega}_i,\sigma},
\label{Eq.306}
\end{equation}
where {$u^M_{\hat{\omega}_i,\sigma}$} (positive frequency, particles) and {$v^M_{\hat{\omega}_i,\sigma}$} (negative frequency, antiparticles) are monochromatic solutions of massless Dirac equation in Minkowski spacetime, and the subscript {$\sigma$} denotes spin. Coefficients {$D_{ij}$} and {$E_{ij}$}  are specifically chosen so that the annihilation and creation operators associated with the constructed Minkowski modes {$\psi^M_{\omega,\sigma}$} and {$\bar{\psi}^M_{\omega,\sigma}$} relate to the Rindler operators via the following Bogoliubov transformation \cite{EMM1}
\begin{equation}
\begin{split}
&c_{\omega,\sigma} = \cos r_{d,i} c_{I,\omega,\sigma} - \sin r_{d,i} d_{IV,\omega,-\sigma}^\dagger,\\
&d^{\dagger}_{\omega,\sigma} = \cos r_{d,i} \, d^{\dagger}_{IV,\omega,\sigma} + \sin r_{d,i} \, c_{I,\omega,-\sigma},
\label{Eq.307}
\end{split}
\end{equation}
where {$c_{\omega,\sigma}$} (particle annihilation) and {$d_{\omega,\sigma}$} (antiparticle annihilation) correspond to {$\psi^M_{\omega,\sigma}$} and {$\bar{\psi}^M_{\omega,\sigma}$} respectively, and $r_{d,i}$ represents the Bogoliubov angle. For Dirac fields, Bogoliubov coefficients take a trigonometric form \cite{PMA,WGU,RJ,ZJ}
\begin{equation}
\alpha_{ij} = \cos r_{d,i} \delta_{ij}, \quad \beta_{ij} = -\sin r_{d,i} \delta_{ij}.
\label{Eq.308}
\end{equation}
Due to the Unruh effect, the Rindler observer perceives the Minkowski vacuum as a thermal state with temperature ${T_U = \frac{a_0}{2\pi}}$. For Dirac fields, the excitation probability follows the Fermi-Dirac distribution
\begin{equation}
n_F(\omega) = \frac{1}{e^{\omega/T_U} + 1}.
\label{Eq.309}
\end{equation}
The squared modulus of the Bogoliubov coefficient $\beta_{ij}$ corresponds precisely to the particle excitation probability in the Rindler vacuum. By using the trigonometric identity $\tan^2(x) = \frac{\sin^2(x)}{1-\sin^2(x)}$, a nontrivial relationship can be obtained as
\begin{equation}
\tan {r_{d,i}} = e^{-\pi\omega/a_0}.
\label{Eq.310}
\end{equation}
For the fixed observer (Bob) staying at a position {$r=r_0$} near the black hole horizon ({$r_0 \approx 2M$}), the acceleration is approximately equal to {$a_0 \approx \frac{\kappa}{\sqrt{1 - \frac{2M}{r_0}}}$}. \cite{EMM1} Then, we have
\begin{equation}
\tan q_{d,i} = e^{ - \frac{\Omega }{2}\sqrt{1 - \frac{1}{R_0}} }.
\label{Eq.311}
\end{equation}
This approximation holds if {$R_0 - 1 \ll 1$}, where {$R_0 =  {r_0}/{2M}$} and {$\Omega = {\omega}/{T_H} = 8\pi\omega M $}.

We expressed the Hartle-Hawking vacuum {$|0_{\omega_i}\rangle_H$} in the Boulware basis \cite{EMM1,EMM2}
\begin{equation}
\begin{split}
\left| {0_{{\omega _i}}} \right\rangle _H &= \left( {\cos {q_{d,i}}} \right)^2\left| {0_{{\omega _i}}} \right\rangle _B\left| {0_{{\omega _i}}} \right\rangle _{\overline B } \\
&+ \sin {q_{d,i}}\cos {q_{d,i}}\left( \left| {{ \uparrow _{{\omega _i}}}} \right\rangle _B\left| {{ \downarrow _{{\omega _i}}}} \right\rangle _{\overline B } + \left| {{ \downarrow _{{\omega _i}}}} \right\rangle _B\left| {{ \uparrow _{{\omega _i}}}} \right\rangle _{\overline B } \right) \\
&+ \left( {\sin {q_{d,i}}} \right)^2\left| {{p_{{\omega _i}}}} \right\rangle _B\left| {{p_{{\omega _i}}}} \right\rangle _{\overline B },
\end{split}
\label{Eq.312}
\end{equation}
where {${p_{{\omega _i}}}$} stands for a pair of spin states in the mode with frequency {${\omega _i}$}. And one particle state of Hartle–Hawking vacuum can be described as
\begin{equation}
\begin{split}
{\left| {{ \uparrow _{{\omega _i}}}} \right\rangle _H} = \cos {q_{d,i}}{\left| {{ \uparrow _{{\omega _i}}}} \right\rangle _B}{\left| {{0_{{\omega _i}}}} \right\rangle _{\overline B }} + \sin {q_{d,i}}{\left| {{p_{{\omega _i}}}} \right\rangle _B}{\left| {{ \uparrow _{{\omega _i}}}} \right\rangle _{\overline B }},
\\
{\left| {{ \downarrow _{{\omega _i}}}} \right\rangle _H} = \cos {q_{d,i}}{\left| {{ \downarrow _{{\omega _i}}}} \right\rangle _B}{\left| {{0_{{\omega _i}}}} \right\rangle _{\overline B }} - \sin {q_{d,i}}{\left| {{p_{{\omega _i}}}} \right\rangle _B}{\left| {{ \downarrow _{{\omega _i}}}} \right\rangle _{\overline B }}.
\label{Eq.313}
\end{split}
\end{equation}
\begin{figure}
\centering 
  \centering
  \includegraphics[width=.50\textwidth]{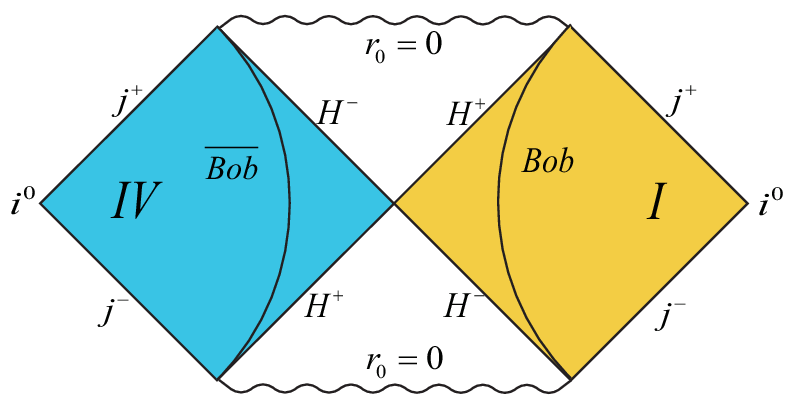}
\caption{The Penrose diagram shows the trajectories for Bob and Anti-Bob. ${i^0}$ denotes the spatial infinities, ${i^-}$ and ${i^+}$ represents the timelike past and future infinities respectively; ${j^-}$ and ${j^+}$ respectively represents the null past and future infinities and ${H^ \pm }$ amount to $r_{0} = 2M$ and denote the event horizons } 
  \label{fig:1}
\end{figure}
For clarity, the Penrose diagram of a Schwarzschild black hole has been offered in figure \ref{fig:1}, where ${\rm I}$ refers to the physically accessible region, while ${\rm IV}$ refers to the physically inaccessible region. For spinless state, the Hartle-Hawking vacuum and its first excitation can be expressed  as \cite{PMA2}
\begin{align}
{\left| {{0_{{\omega _i}}}} \right\rangle _\mathrm{H}} &= {\left[ {1 + {e^{\left( { - \Omega \sqrt {1 - {1 \mathord{\left/{\vphantom {1 {{R_0}}}} \right.\kern-\nulldelimiterspace} {{R_0}}}} } \right)}}} \right]^{ - \frac{1}{2}}}{\left| {{0_{{\omega _i}}}} \right\rangle _\mathrm{I}}{\left| {{0_{{\omega _i}}}} \right\rangle _\mathrm{IV}} \nonumber \\
\label{Eq.314}
&\quad + {\left[ {1 + {e^{\left( { \Omega \sqrt {1 - {1 \mathord{\left/{\vphantom {1 {{R_0}}}} \right.\kern-\nulldelimiterspace} {{R_0}}}} } \right)}}} \right]^{ - \frac{1}{2}}}{\left| {{1_{{\omega _i}}}} \right\rangle _\mathrm{I}}{\left| {{1_{{\omega _i}}}} \right\rangle _\mathrm{IV}}, \\
{\left| {{1_{{\omega _i}}}} \right\rangle _\mathrm{H}} &= {\left| {{1_{{\omega _i}}}} \right\rangle _\mathrm{I}}{\left| {{0_{{\omega _i}}}} \right\rangle _\mathrm{IV}}.
\label{Eq.315}
\end{align}
The subscripts I and IV stand for the outside and inside of the horizon.

\section{Quantum properties under Schwarzschild black hole}
\label{sec4}
In this section, we focus on examining quantumness in the context of Schwarzschild black hole, including quantum coherence, quantum entanglement, and uncertainty relation. Interestingly, we find that the coherence and entanglement of the multipartite states are identical in the current consideration. Moreover, it turns out that the proposed uncertainty relation is tighter than the previous one presented by Rene and Boileau \cite{JMR}, and interestingly find the uncertainty bound is anti-correlated with the coherence.

\subsection{$N$-partite coherence in Schwarzschild spacetime}
\label{sec4.1}
Quantum coherence is an important quantum resource, which plays a significant role in quantum computing and quantum communication. Quantification of quantum coherence was first proposed by Baumgratz \textit{et al.} from the perspective of resource theory, introducing two standard measures: the relative entropy of coherence (REC) and the $l_1$-norm of coherence. The $l_1$-norm is computationally and analytically more tractable than the REC, especially in examining the physical meaning of coherence within multipartite systems. In this subsection, we use ${l_1}$-norm to measure coherence of the multipartite systems. For an $N$-particle system with a density matrix of the X state, the density matrix can be expressed as
\begin{equation}
\begin{aligned}
\hat{\rho}_X &= \begin{pmatrix}
a_1 & & & & & & & c_1 \\
& a_2 & & & & & c_2 & \\
& & \ddots & & & \vdots & & \\
& & & a_n & c_n & & & \\
& & & c_n^* & b_n & & & \\
& & \vdots & & & \ddots & & \\
& c_2^* & & & & & b_2 & \\
c_1^* & & & & & & & b_1
\end{pmatrix} 
= \begin{pmatrix}
A & C \\
C^T & B
\end{pmatrix},
\end{aligned}
\label{Eq.401}
\end{equation}\\
where $n = {2^{N - 1}}$. In the density matrix $\hat{\rho}_X$, $A$, $B$, and $C$ are the corresponding sub-matrices with elements $a_i$, $b_j$, and $c_k$ respectively, and $C^T$ is the conjugate of $C$.\\
In this article, we use the $l_1$-norm to measure the coherent properties of the Schwarzschild black hole. The ${l_1}$-norm computes the sum of the absolute values of the nondiagonal elements of the density matrix:
\begin{equation}
{C_{{l_1}}}(\hat \rho ) = \sum\nolimits_{i \ne j} {\left| {{\hat \rho _{i,j}}} \right|}. 
\label{Eq.402}
\end{equation}
Therefore for the density matrix of the X state, the ${l_1}$-norm depends on the value of ${c_i}$ and ${c_i^*}$.

Initially, we assume that $N$ particles share a GHZ-type state and locate at asymptotically flat region of Schwarzschild black hole
\begin{equation}
\left| \phi  \right\rangle _{1,...,N}^{\rm GHZ} = z\left| {{0_1}{0_2} \cdots {0_N}} \right\rangle  + \sqrt {1 - {z^2}} \left| {{1_1}{1_2} \cdots {1_N}} \right\rangle,
\label{Eq.403}
\end{equation}
where the state parameter ${z \in [0,1]}$. Then, we let $n$ observers hover near the event horizon and the rest $N-n$ observers remain at asymptotically flat region. For the $n$ particles near the event horizon of a black hole, due to the Hawking radiation effect, a two-mode squeezing effect will occur. $n$ Kruskal modes are mapped into $2n$ Schwarzschild modes ($n$ modes represent the interior region of the black hole, while $n$ modes represent the exterior region of the black hole). So we can rewrite  Eq. (\ref{Eq.403}) by using evolution eqs (\ref{Eq.314}) and (\ref{Eq.315}).
\begin{equation}
\begin{split}
\left| \phi  \right\rangle _{1,...,N + n}^{\rm GHZ} &= z[({\left| 0 \right\rangle _{n + 1}}{\left| 0 \right\rangle _{n + 2}} \cdots {\left| 0 \right\rangle _N}) \\
&\quad \mathop  \otimes \limits_{i = 1}^n (\frac{1}{{\sqrt {{e^{ - \Omega \sqrt {1 - {1 \mathord{\left/
 {\vphantom {1 {{R_0}}}} \right.
 \kern-\nulldelimiterspace} {{R_0}}}} }} + 1} }}{\left| 0 \right\rangle _{{\rm I},i}}{\left| 0 \right\rangle _{{\rm IV},i}}\\
 &\quad + \frac{1}{{\sqrt {{e^{\Omega \sqrt {1 - {1 \mathord{\left/
 {\vphantom {1 {{R_0}}}} \right.
 \kern-\nulldelimiterspace} {{R_0}}}} }} + 1} }}{\left| 1 \right\rangle _{{\rm I},i}}{\left| 1 \right\rangle _{{\rm IV},i}})]\\
&\quad  + \sqrt {1 - {z^2}} [({\left| 1 \right\rangle _{n + 1}}{\left| 1 \right\rangle _{n + 2}} \cdots {\left| 1 \right\rangle _N})\\
&\quad \mathop  \otimes \limits_{i = 1}^n ({\left| 1 \right\rangle _{{\rm I},i}}{\left| 0 \right\rangle _{{\rm IV},i}})].
\label{Eq.404}
\end{split}
\end{equation}
Suppose that $p$ of the $n$ particles near the event horizon are in the physically accessible region and $q$ of the particles are in the physically inaccessible region $(p+q=n)$. We take trace of physically accessible $p$ particles and physically inaccessible $q$ particles of $\left| \phi  \right\rangle _{1,...,N + n}^{\rm GHZ}$ and we let $\left| {\bar 0} \right\rangle  = {\left| 0 \right\rangle _{n + 1}}{\left| 0 \right\rangle _{n + 2}} \cdots {\left| 0 \right\rangle _N}$ and $\left| {\bar 1} \right\rangle  = {\left| 1 \right\rangle _{n + 1}}{\left| 1 \right\rangle _{n + 2}} \cdots {\left| 1 \right\rangle _N}$. Inspired by the derivation of quantum systems under dilaton black holes \cite{WML,SMW2}, we can obtain the system ${\hat \rho _{N - n,p,q}}$ in Schwarzschild spacetime
\begin{widetext}
{\footnotesize
\begin{align}
\hat{\rho}&_{N-n,p,q} = z^2 \xi^{2n} \left| \bar{0} \right\rangle \left\langle \bar{0} \right| \left[ \bigotimes_{i=1}^p \left( \left| 0 \right\rangle_{\mathrm{I},i} \left\langle 0 \right| \right) \right] \left[ \bigotimes_{j=1}^q \left( \left| 0 \right\rangle_{\mathrm{IV},j} \left\langle 0 \right| \right) \right] \nonumber  \\
& + z^2 \xi^{2n-2} \eta^2 \left| \bar{0} \right\rangle \left\langle \bar{0} \right| \times \left\{\sum\limits_{m = 1}^p {\left[ {\left( {{{\left| 1 \right\rangle }_{{\rm{I}},m}}\langle 1|} \right)} \right.} \mathop  \otimes \limits_{i = 1,i \ne m}^p \left( {{{\left| 0 \right\rangle }_{{\rm{IV}},m}}\langle 0|} \right)\mathop  \otimes \limits_{j = 1}^q \left( {{{\left| 0 \right\rangle }_{{\rm{I}},j}}\langle 0|} \right) + \sum_{m=1}^q \left[ \bigotimes_{i=1}^p \left( \left| 0 \right\rangle_{\mathrm{I},i} \langle 0| \right) 
    \left( \left| 1 \right\rangle_{\mathrm{IV},m} \langle 1| \right) \right. \right. \left. \left. \bigotimes_{j=1,j\ne m}^q \left( \left| 0 \right\rangle_{\mathrm{IV},j} \langle 0| \right) \right] \right\} \nonumber \\
& + z^2 \xi^{2n-4} \eta^4 \left| \bar{0} \right\rangle \left\langle \bar{0} \right| \times \left\{ \sum\limits_{m = 2}^p {\sum\limits_{l = 1}^{m - 1} {\left[ {\left( {{{\left| 1 \right\rangle }_{{\rm{I}},l}}\langle 1|} \right)\left( {{{\left| 1 \right\rangle }_{{\rm{I}},m}}\langle 1|} \right)} \right]} } {\rm{ }}\mathop  \otimes \limits_{i = 1,i \ne l,m}^p \left( {{{\left| 0 \right\rangle }_{{\rm{IV}},i}}\langle 0|} \right)\mathop  \otimes \limits_{j = 1}^q \left( {{{\left| 0 \right\rangle }_{{\rm{I}},j}}\langle 0|} \right)\right. + \sum_{m=2}^q \sum_{l=1}^{m-1} \left[ \bigotimes_{i=1}^p \left( \left| 0 \right\rangle_{\mathrm{I},i} \left\langle 0 \right| \right) 
    \left( \left| 1 \right\rangle_{\mathrm{IV},l} \langle 1| \right) 
    \left( \left| 1 \right\rangle_{\mathrm{IV},m} \langle 1| \right) \right. \nonumber  \\
& \left. \bigotimes_{j=1,j\ne l,m}^q \left( \left| 0 \right\rangle_{\mathrm{IV},j} \langle 0| \right) \right] \left. + \sum_{l=1}^p \sum_{m=1}^q \left[ \left( \left| 1 \right\rangle_{\mathrm{I},l} \langle 1| \right) 
    \bigotimes_{i=1,i\ne l}^p \left( \left| 0 \right\rangle_{\mathrm{I},i} \left\langle 0 \right| \right) 
    \left( \left| 1 \right\rangle_{\mathrm{IV},m} \langle 1| \right) \right. \right.  \left. \left. \bigotimes_{j=1,j\ne m}^q \left( \left| 0 \right\rangle_{\mathrm{IV},j} \langle 0| \right) \right] \right\} \nonumber \\
& + \cdots + \left( \frac{z}{\eta} \right)^2 \left( \left| \bar{0} \right\rangle \left\langle \bar{0} \right| \right) \bigotimes_{i=1}^p \left( \left| 0 \right\rangle_{\mathrm{I},i} \left\langle 0 \right| \right) \bigotimes_{j=1}^q \left( \left| 1 \right\rangle_{\mathrm{IV},j} \langle 1| \right)  + z \sqrt{1-z^2} \xi^p \eta^q \left[ \left( \left| \bar{0} \right\rangle \left\langle \bar{1} \right| \right) 
    \bigotimes_{i=1}^p \left( \left| 0 \right\rangle_{\mathrm{I},i} \left\langle 1 \right| \right) \right. \left. \bigotimes_{j=1}^q \left( \left| 1 \right\rangle_{\mathrm{IV},j} \langle 0| \right) \right] \nonumber \\
&+ (1-z^2) \left( \left| \bar{1} \right\rangle \left\langle \bar{1} \right| \right) \bigotimes_{i=1}^p \left( \left| 0 \right\rangle_{\mathrm{I},i} \left\langle 0 \right| \right) \bigotimes_{j=1}^q \left( \left| 1 \right\rangle_{\mathrm{IV},j} \langle 0| \right).
\label{Eq.405}
    \end{align}
}
\end{widetext}
Here, $\xi  = \frac{1}{{\sqrt {{e^{ - \Omega \sqrt {1 - 1/{R_0}} }} + 1} }}$ and $\eta  = \frac{1}{{\sqrt {{e^{\Omega \sqrt {1 - 1/{R_0}} }} + 1} }}$. Obviously, we find that only the two diagonal elements are not zero. In other words, ${\hat \rho _{N - n,p,q}}$ is a density matrix of X states, so we can represent the matrix as the matrix form
\begin{equation}
{\hat \rho _{N - n,p,q}} = \left( {\begin{array}{*{20}{c}}
A&C\\
{{C^T}}&B
\end{array}} \right).
\label{Eq.406}
\end{equation}
From Eq. (\ref{Eq.405}), we can see that the sub-matrixes $B$ and $C$ are {${2^n} \times {2^n}$ dimensions and are detailed in Appendix \ref{appendixB}. In the sub-matrix $B$, only ${b_{{2^q}}} = 1 - {z^2}$ is not zero and we can see that ${c_{{2^q}}} = \frac{{z\sqrt {1 - {z^2}} }}{{\sqrt {{{({e^{ - \Omega \sqrt {{{1 - 1} \mathord{\left/ {\vphantom {{1 - 1} {{R_0}}}} \right. \kern-\nulldelimiterspace} {{R_0}}}} }} + 1)}^p}{{({e^{\Omega \sqrt {{{1 - 1} \mathord{\left/ {\vphantom {{1 - 1} {{R_0}}}} \right. \kern-\nulldelimiterspace} {{R_0}}}} }} + 1)}^q}} }}$ is not zero. As a result, one can obtain the $l_1$-norm of ${\rho _{N - n,p,q}}$

\begin{equation}
\begin{split}
{C_{{l_1}}}({\hat \rho _{N - n,p,q}})&= {c_{{2^q}}} + c_{{2^q}}^*\\
&  = \frac{{2z\sqrt {1 - {z^2}} }}{{\sqrt {{{({e^{ - \Omega \sqrt {{{1 - 1} \mathord{\left/
 {\vphantom {{1 - 1} {{R_0}}}} \right.
 \kern-\nulldelimiterspace} {{R_0}}}} }} + 1)}^p}{{({e^{\Omega \sqrt {{{1 - 1} \mathord{\left/
 {\vphantom {{1 - 1} {{R_0}}}} \right.
 \kern-\nulldelimiterspace} {{R_0}}}} }} + 1)}^q}} }}.
\label{Eq.407}
\end{split}
\end{equation}
\begin{figure*}[htbp]
\subfigure[]
  {
      \label{fig:2a}
      \includegraphics[width=5.5cm]{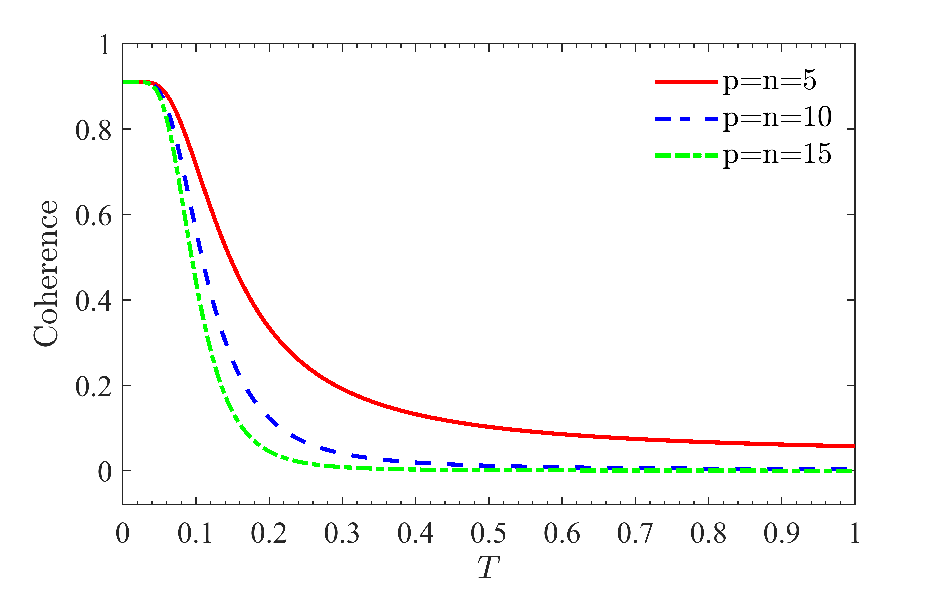}
  }
\subfigure[]
  {
      \label{fig:2b}
      \includegraphics[width=5.5cm]{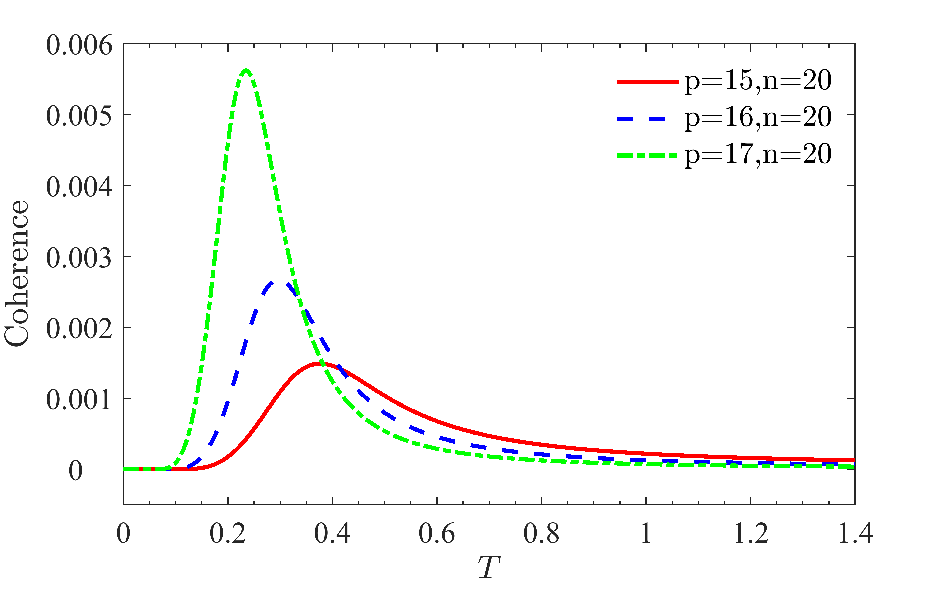}
  }
\subfigure[]
  {
      \label{fig:2c}
      \includegraphics[width=5.5cm]{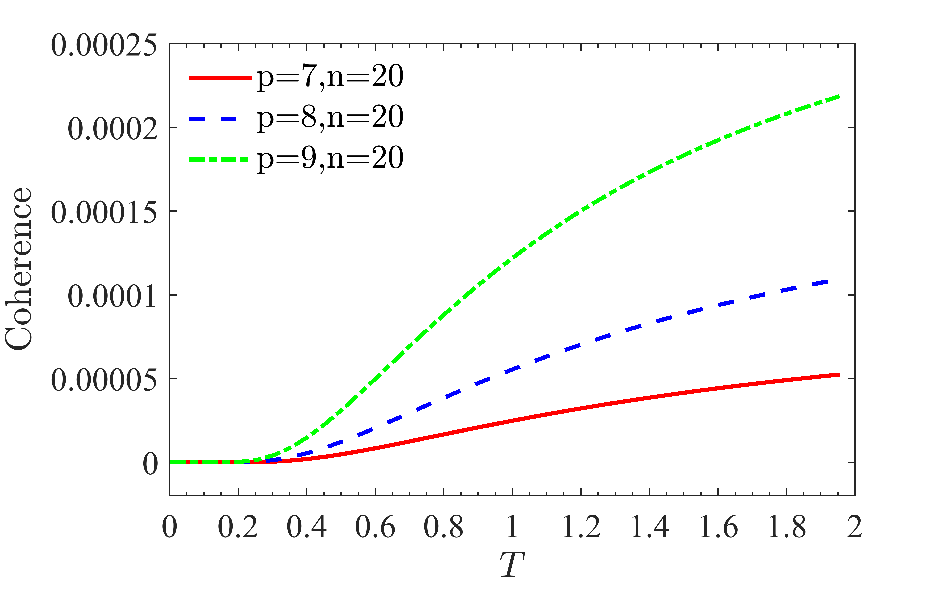}
  }

\subfigure[]
  {
      \label{fig:2d}
      \includegraphics[width=5.5cm]{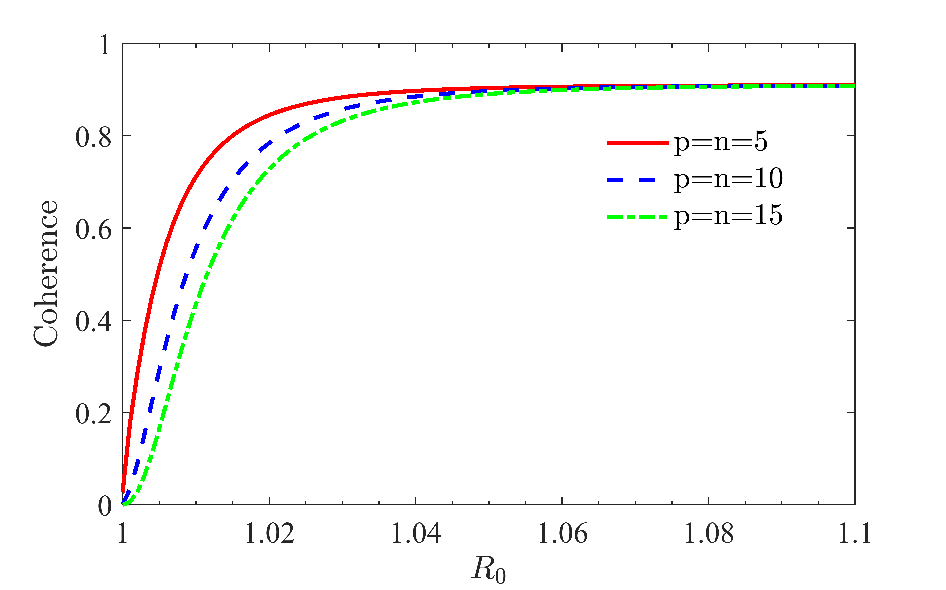}
  }
\subfigure[]
  {
      \label{fig:2e}
      \includegraphics[width=5.5cm]{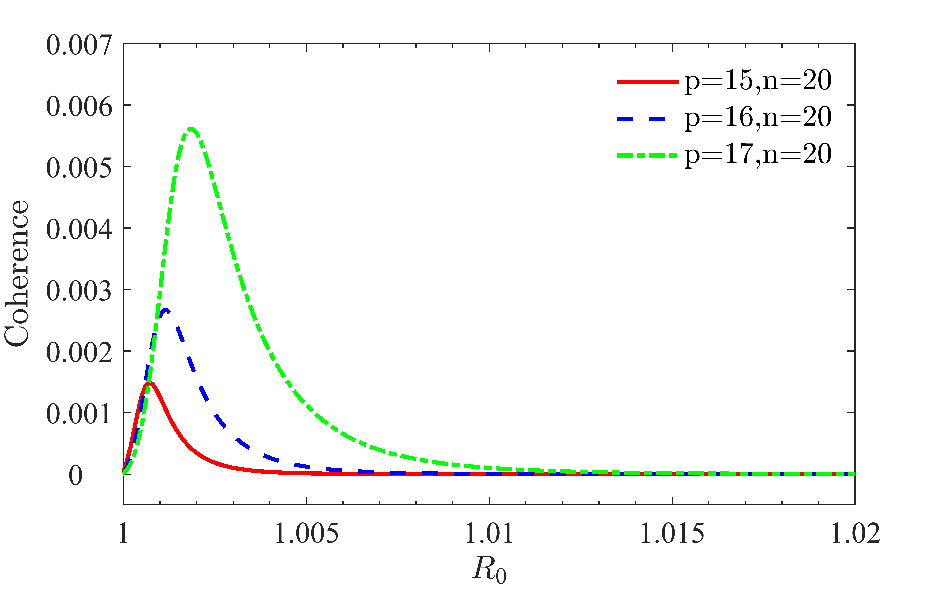}
  }
\subfigure[]
  {
      \label{fig:2f}
      \includegraphics[width=5.5cm]{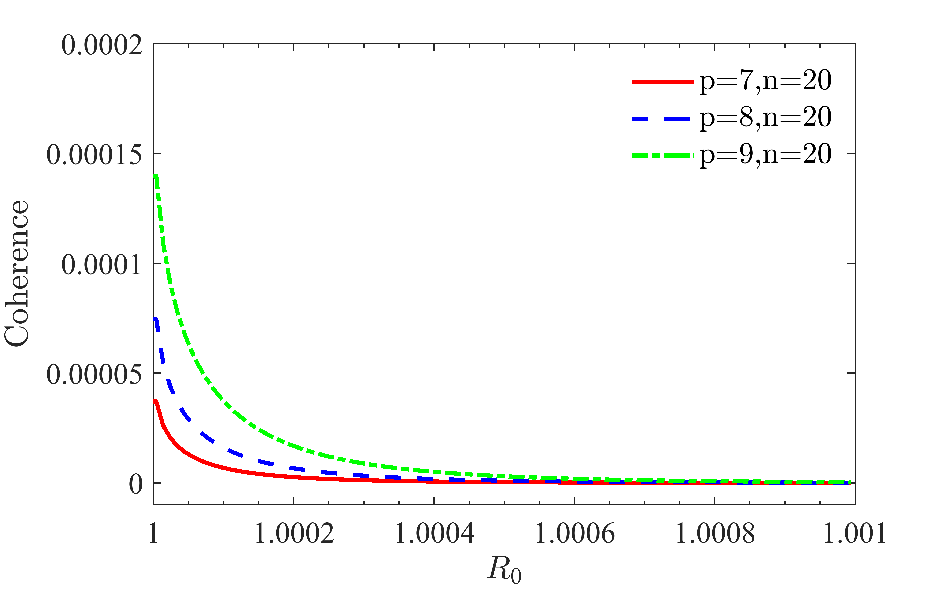}
  }
\caption{$l_1$-norm ${C_{{l_1}}}({\hat \rho _{N - n,p,q}})$ as functions of $T$ and $R_0$ for different $p$, $q$ and $n$. Graphs (a) and (d) correspond to the case where $p = n$ and $q = 0$. Graphs (b) and (e) represent the case where $p > q$ and graphs (c) and (f) represent the case of $p < n$ respectively.}
\label{fig:2}
\end{figure*}
It is well-known that genuine $N$-partite entanglement is an important quantum resource \cite{HJB,ASS,MMD,VS}. To quantify this resource, we employ the genuine multipartite (GM) concurrence \cite{SMH}, which is a direct generalization of the Wootters' concurrence for $N$-partite X-state density matrix \cite{WKW}. The mixed state GM concurrence is defined as \cite{ZHM}
\begin{equation}
\begin{split}
C_{\text{GM}}(\hat{\rho}) = \min_{\alpha} \sum_i p_i^{(\alpha)} C_{\text{GM}}(|\psi_i^{(\alpha)}\rangle),
\label{Eq.408}
\end{split}
\end{equation}
where the minimization is taken over all pure-state decompositions $\hat{\rho} = \sum_i p_i^{(\alpha)} |\psi_i^{(\alpha)}\rangle\langle\psi_i^{(\alpha)}|$, and $C_{\text{GM}}(|\psi\rangle)$ for pure states is defined as:
\begin{equation}
\begin{split}
C_{\text{GM}}(|\psi\rangle) = \min_{j} \sqrt{2\left(1-\Pi_j(|\psi\rangle)\right)},
\label{Eq.409}
\end{split}
\end{equation}
where $\Pi_j(|\psi\rangle)$ denotes the purity with respect to bipartition $j$. One can consider the orthonormal basis of $N$ qubits matrix given by the computational basis: ${\{|0,0,\ldots,0\rangle, |0,0,\ldots,1\rangle, \ldots, |1,1,\ldots,1\rangle\}}$. In this basis, the GM concurrence of an $N$-qubit X state is given by \cite{SMH}

\begin{equation}
\begin{split}
{C_\mu }({\hat \rho _X}) &= 2\max \{ 0, |{c_i}| - \sum_{j \ne i}^{2^n} \sqrt{a_j b_j} \} = 2\max \{ 0, |{c_{2^q}}| \} \\
&= \frac{{2z\sqrt {1 - {z^2}} }}{{\sqrt {{{({e^{ - \Omega \sqrt {{{1 - 1} \mathord{\left/
 {\vphantom {{1 - 1} {{R_0}}}} \right.
 \kern-\nulldelimiterspace} {{R_0}}}} }} + 1)}^p}{{({e^{\Omega \sqrt {{{1 - 1} \mathord{\left/
 {\vphantom {{1 - 1} {{R_0}}}} \right.
 \kern-\nulldelimiterspace} {{R_0}}}} }} + 1)}^q}} }}\\
 &= {C_{{l_1}}}({\hat \rho _{N - n,p,q}}). 
\label{Eq.408}
\end{split}
\end{equation}
Surprisingly, it is found that the $N$-particle GHZ-type concurrence and $l_1$-norm are identical in the current framework of the black hole.

Figure \ref{fig:2} shows the behavior of $N$-particle GHZ-type state coherence in Schwarzschild black holes under different distance and Hawking temperature. As shown in figure \ref{fig:2}, we can conclude: (i) when $p=n$, the physically accessible coherence decreases monotonically with temperature, increases monotonically with distance, and then both reach fixed values; (ii) when $p<n$, the physically inaccessible coherence exhibit an opposite trend to the physically achievable coherence as the functions of distance and temperature; (iii) when $p>n$, the physically inaccessible coherence does not change linearly, and the peak position will shift as the relative value of $p$ changes.

\subsection{Generalized entropic uncertainty relation  in Schwarzschild spacetime}
\label{sec4.2}
\begin{figure*}[htbp]
\subfigure[]{
  \label{fig:3a}
  \includegraphics[width=0.45\textwidth]{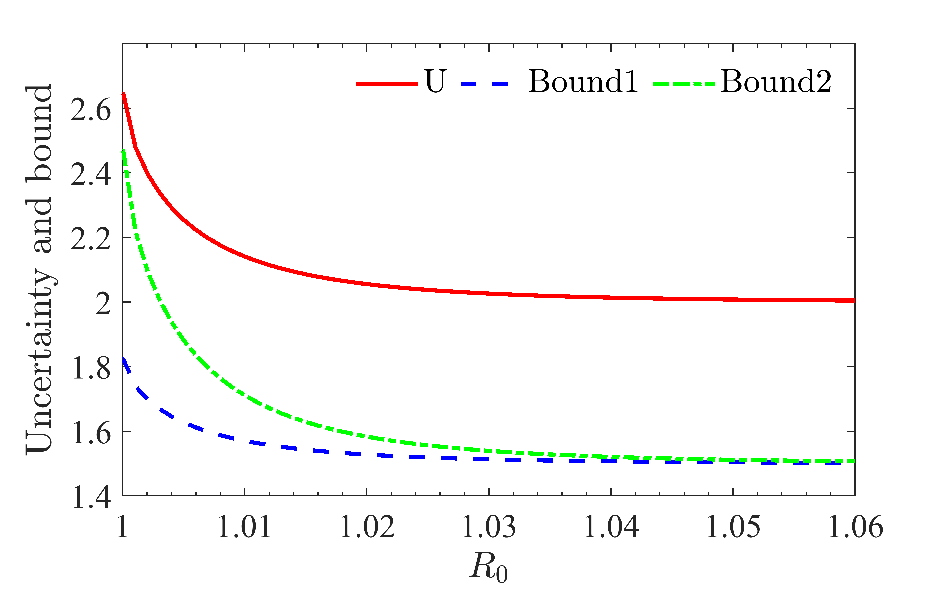}
}
\subfigure[]{
  \label{fig:3b}
  \includegraphics[width=0.45\textwidth]{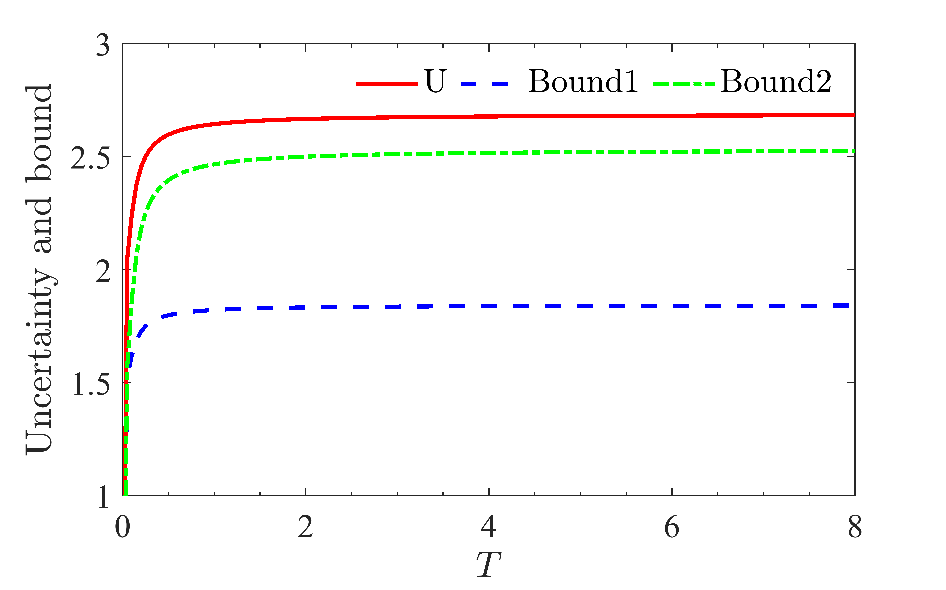}
}\\
\subfigure[]{
  \label{fig:3c}
  \includegraphics[width=0.45\textwidth]{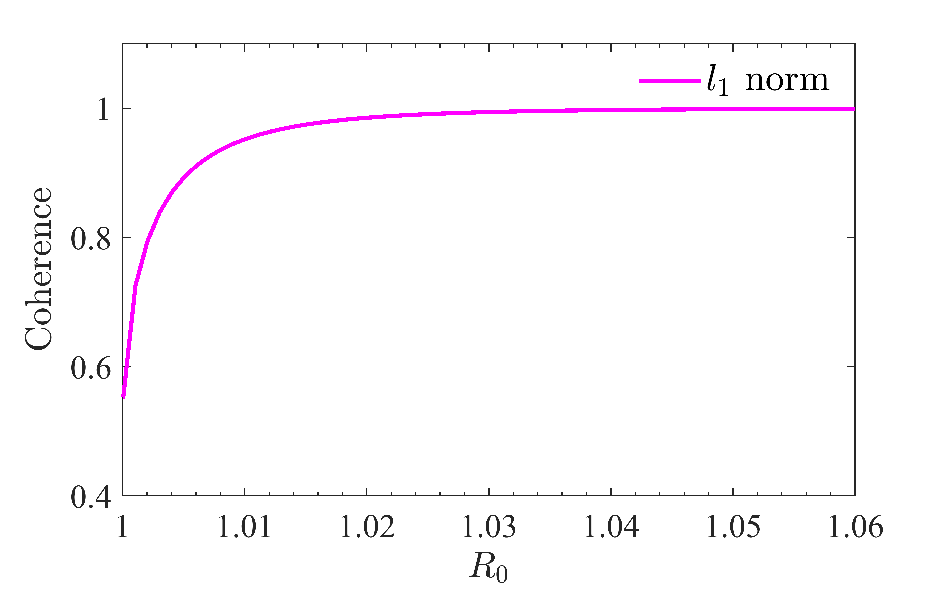}
}
\subfigure[]{
  \label{fig:3d}
  \includegraphics[width=0.45\textwidth]{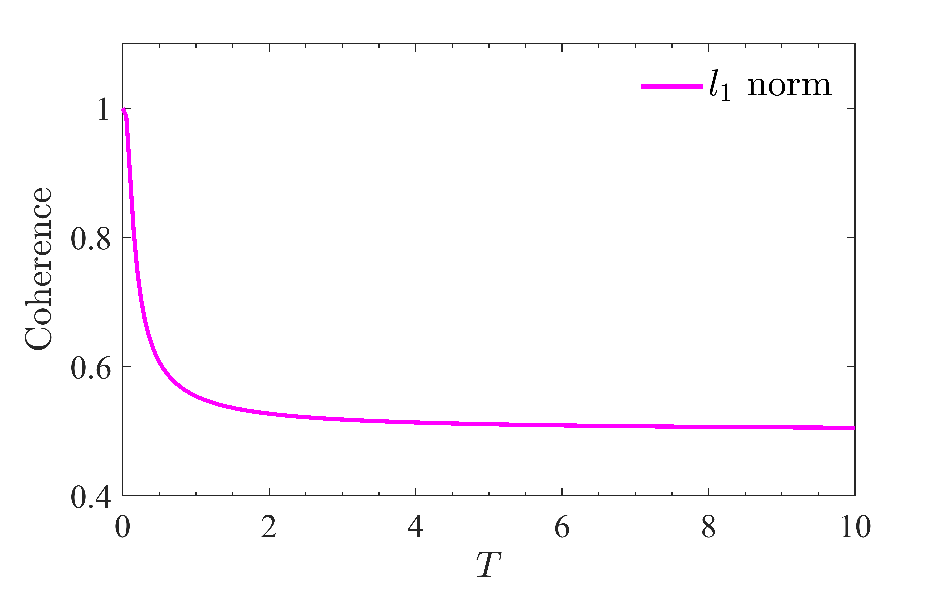}
}
\caption{Graph (a) shows the uncertainty and two bounds with distance $R_0$ for $\Omega = \omega_k / T = 30$ and $z = \frac{\sqrt{2}}{2}$. Graph (b) describes the uncertainty and two bounds with Hawking Temperature $T$ with ${\omega _k} = 1$, ${{R_0} = r \mathord{\left/ {\vphantom {r {{R_H} = 1.05}}} \right. \kern-\nulldelimiterspace} {{R_H} = 1.05}}$ and $z = \frac{{\sqrt 2 }}{2}$. Here, $U$ represents the entropic uncertainty, Bound1 is the lower bound proposed by Renes and Boileau \cite{JMR}, and Bound2 is our proposed new lower bound. Graphs (c) and (d) plot the $l_1$-norm as functions of $R_0$ and $T$ respectively.}
\label{fig:3}
\end{figure*}
\begin{figure*}[htbp]
\subfigure[]{
  \label{fig:4a}
  \includegraphics[width=0.45\textwidth]{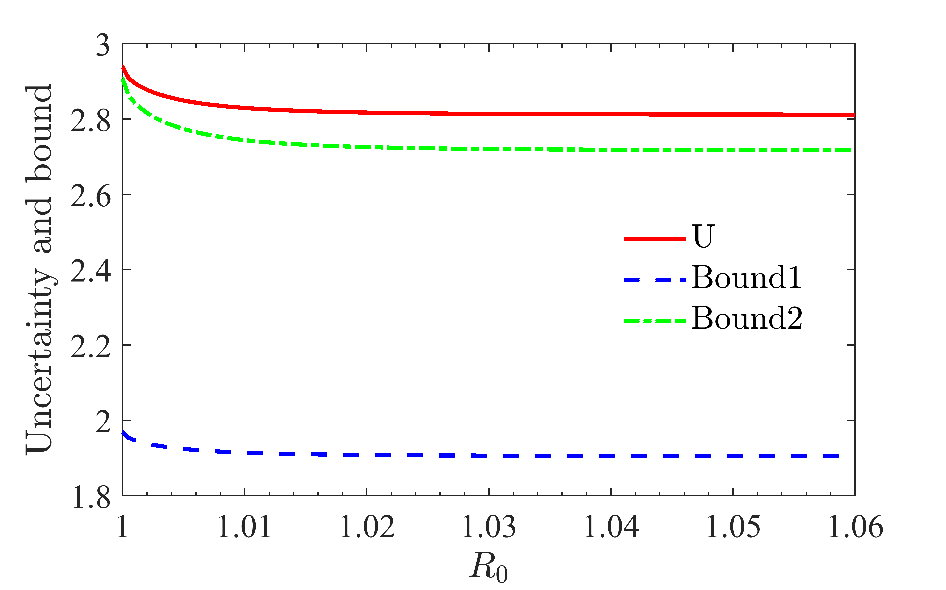}
}
\subfigure[]{
  \label{fig:4b}
  \includegraphics[width=0.45\textwidth]{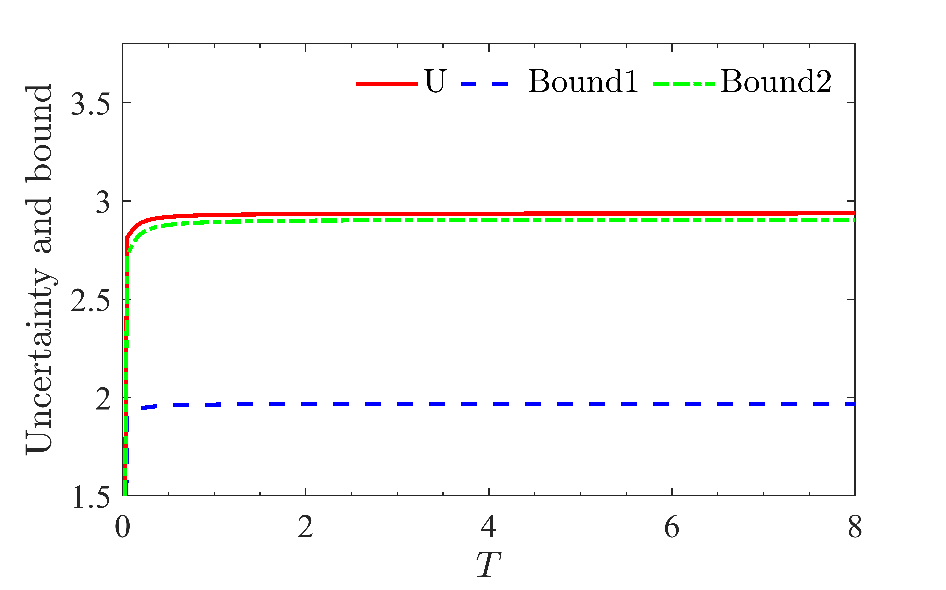}
}\\
\subfigure[]{
  \label{fig:4c}
  \includegraphics[width=0.45\textwidth]{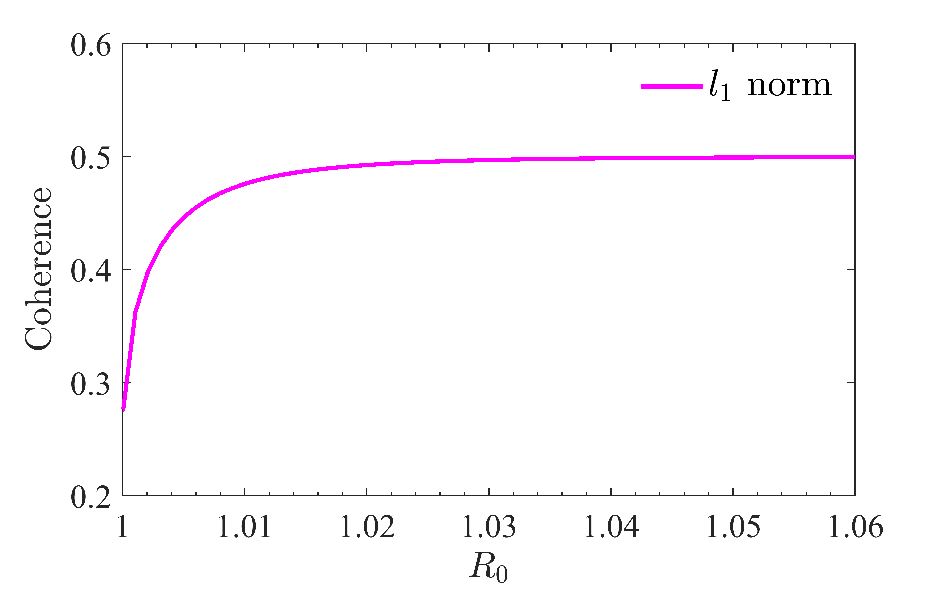}
}
\subfigure[]{
  \label{fig:4d}
  \includegraphics[width=0.45\textwidth]{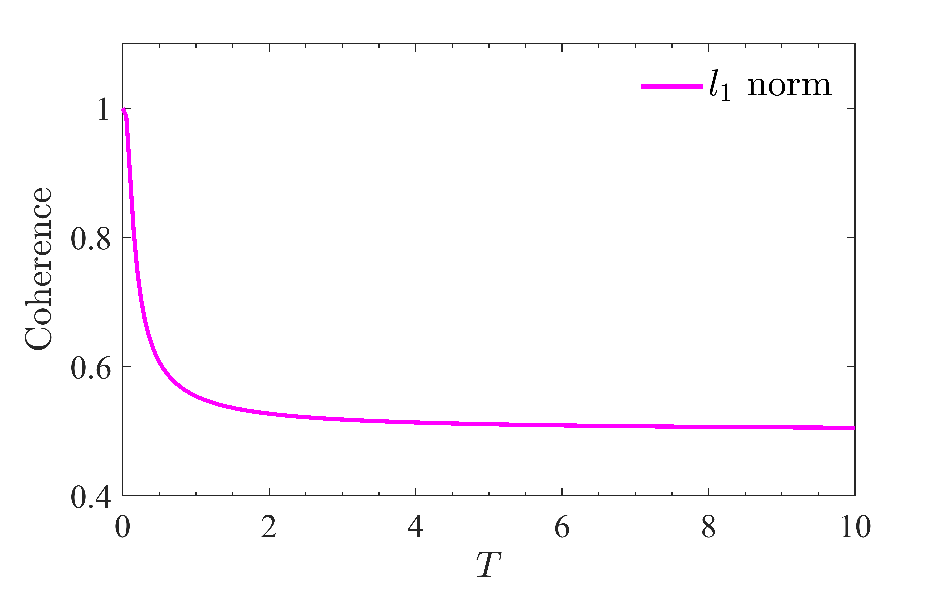}
}
\caption{Entropic uncertainty and coherence with respect to distance $r$ and Hawking Temperature $T$ of Werner state $\hat \rho _{A{B\rm{_I}}{C\rm{_I}}}^{\rm Werner}$ in Schwarzschild black hole. The system purity $p = 0.5$ and state parameter $z = \frac{{\sqrt 2 }}{2}$. Graph (a) and Graph (b) describe the relationship of EURs with distance and temperature respectively. In these graphs, U, Bound1, and Bound2 denote the entropic uncertainty, the lower bound proposed by Renes, and our newly proposed lower bound respectively. Graph (a) presents the EUR vs distance $R_0$ for $\Omega = \omega_k / T = 30$ and $z = \frac{\sqrt{2}}{2}$. Graph (b) describes the EUR with Hawking Temperature $T$ with ${\omega _k} = 1$, ${{R_0} = r \mathord{\left/ {\vphantom {r {{R_H} = 1.05}}} \right. \kern-\nulldelimiterspace} {{R_H} = 1.05}}$ and $z = \frac{{\sqrt 2 }}{2}$. Graph (c) and (d) present the $l_1$-norm as the function of $R_0$ and $T$ respectively.}
\label{fig:4}
\end{figure*}
\quad 
\begin{figure*}[htbp]
\subfigure[]
  {
      \label{fig:5a}
      \includegraphics[width=5.5cm]{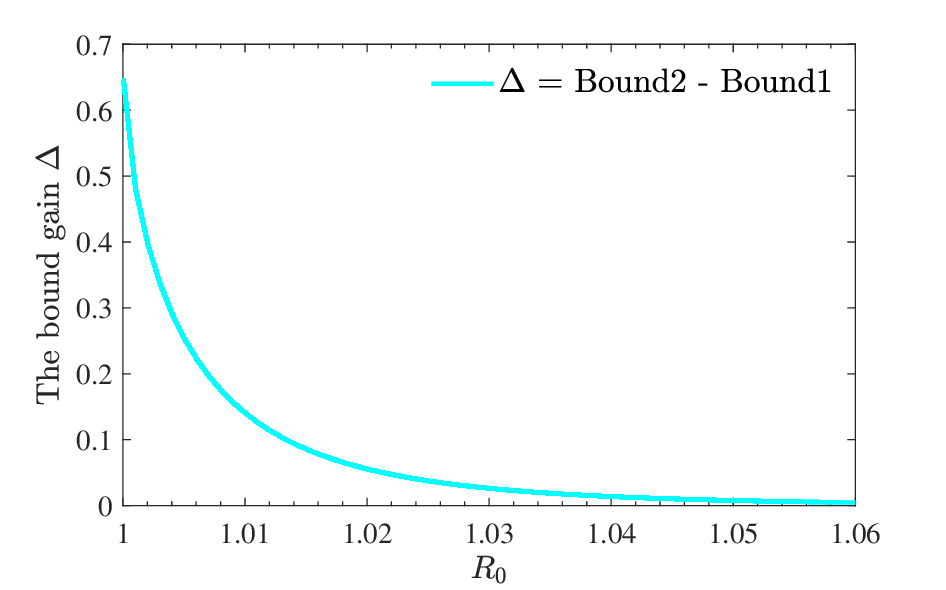}
  }
\subfigure[]
  {
      \label{fig:5b}
      \includegraphics[width=5.5cm]{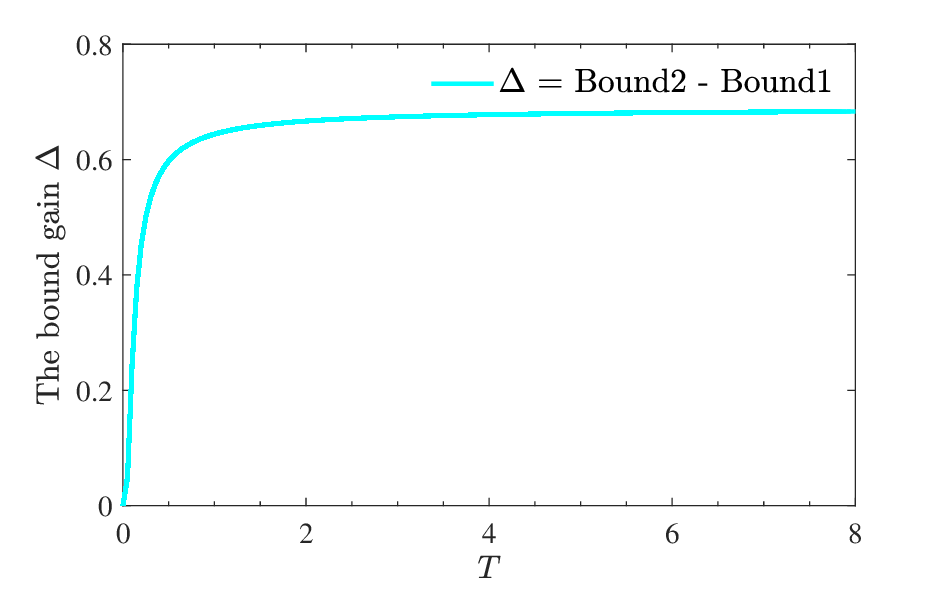}
  }
\subfigure[]
  {
      \label{fig:5c}
      \includegraphics[width=5.5cm]{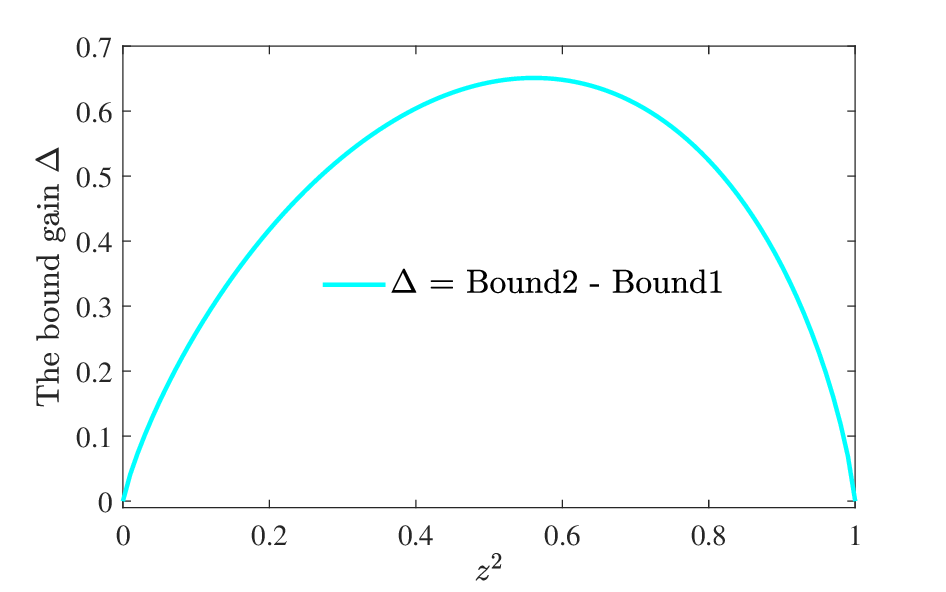}
  }

\subfigure[]
  {
      \label{fig:5d}
      \includegraphics[width=5.5cm]{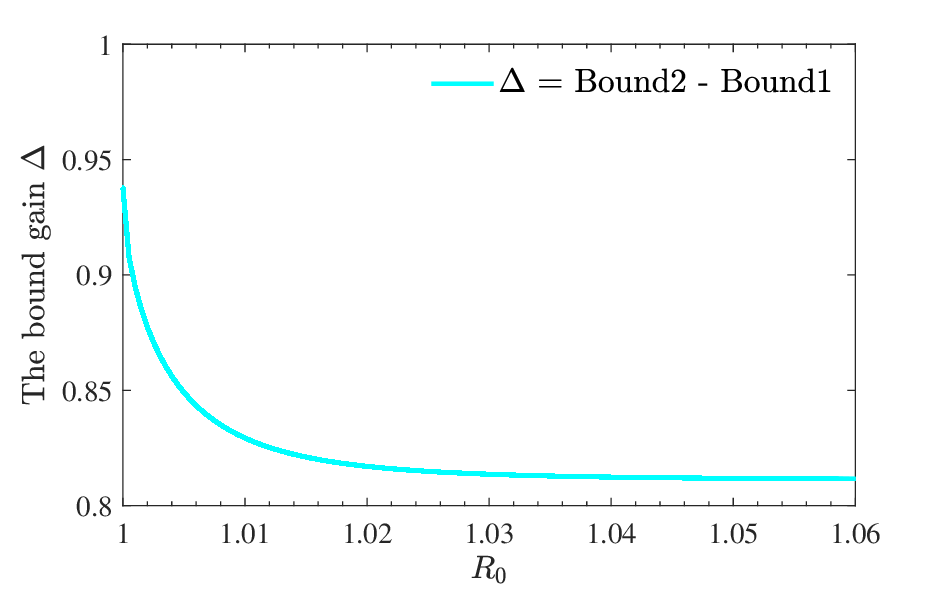}
  }
\subfigure[]
  {
      \label{fig:5e}
      \includegraphics[width=5.5cm]{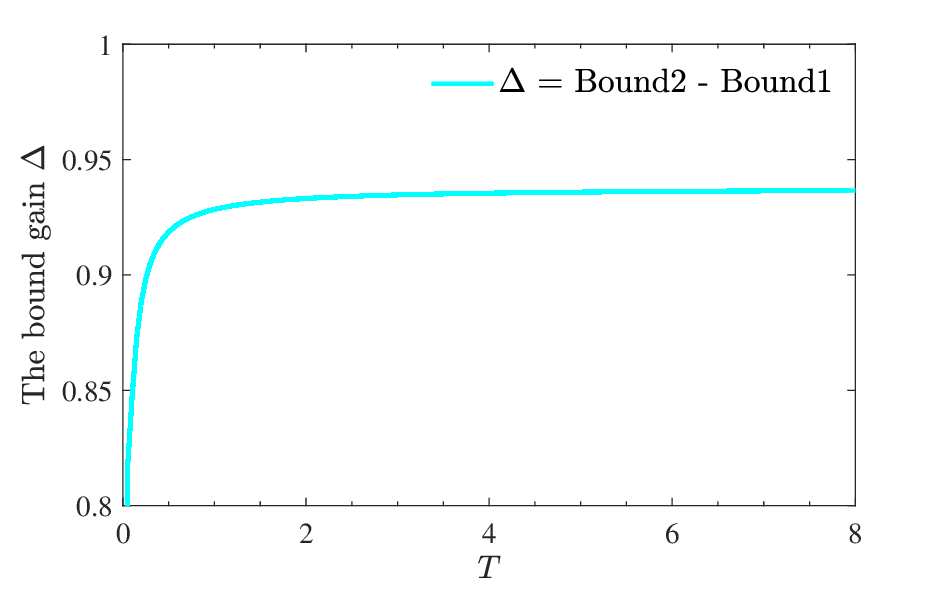}
  }
\subfigure[]
  {
      \label{fig:5f}
      \includegraphics[width=5.5cm]{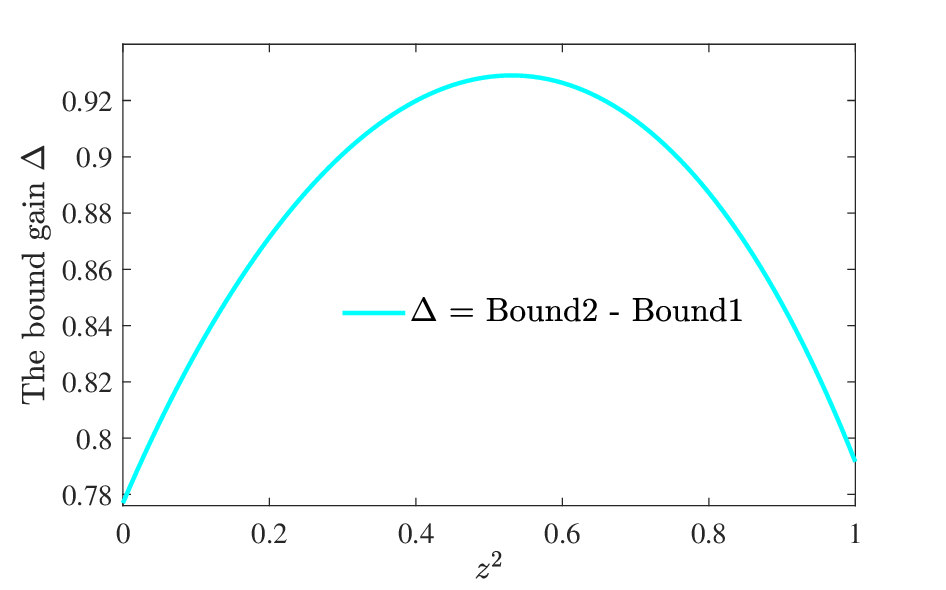}
  }
\caption{The bound gain ($\Delta={\rm Bound}2-{\rm Bound}1$) when the initial states are chosen to GHZ-type and Werner states in the Schwarzschild black hole. Graphs (a), (b) and (c) show the bound gain versus $R_0$, $T$ and $z$ of GHZ-type states, while Graphs (d), (e) and (f) represent the the Werner states respectively. Graphs (a) and (d): $\Delta$ vs $R_0$ with $\Omega = 30$ and ${z^2} = \frac{1}{2}$; Graphs (b) and (e): $\Delta$ vs $T$ with $R_0 = 1.05$ and ${z^2} = \frac{1}{2}$;  Graphs (c) and (f): $\Delta$ vs $z$ with $\Omega = 1$ and $R_0 = 1.05$.}
\label{fig:5}
\end{figure*}
\begin{figure*}[htbp]
\subfigure[]{
  \label{fig:6a}
  \includegraphics[width=0.45\textwidth]{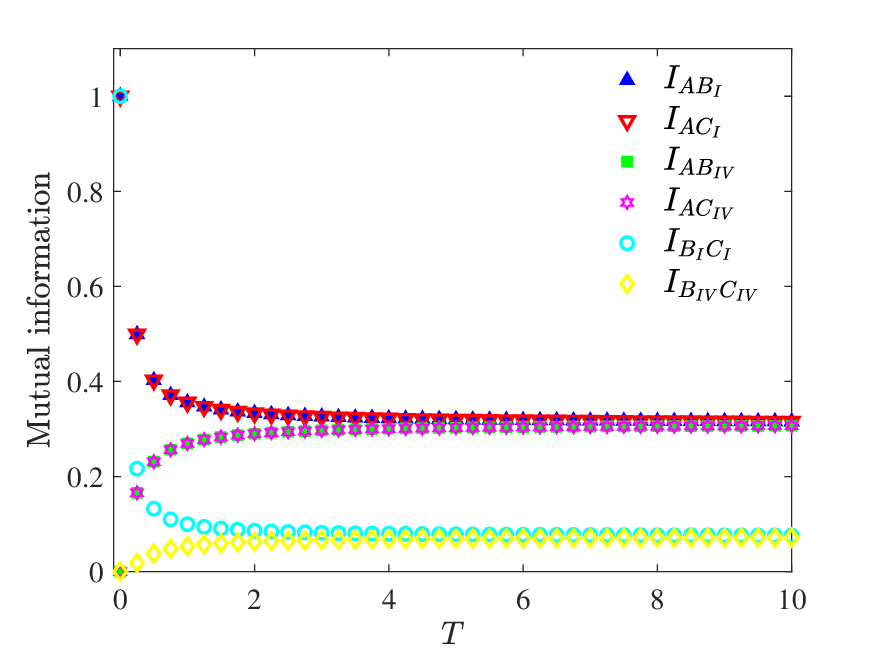}
}
\subfigure[]{
  \label{fig:6b}
  \includegraphics[width=0.45\textwidth]{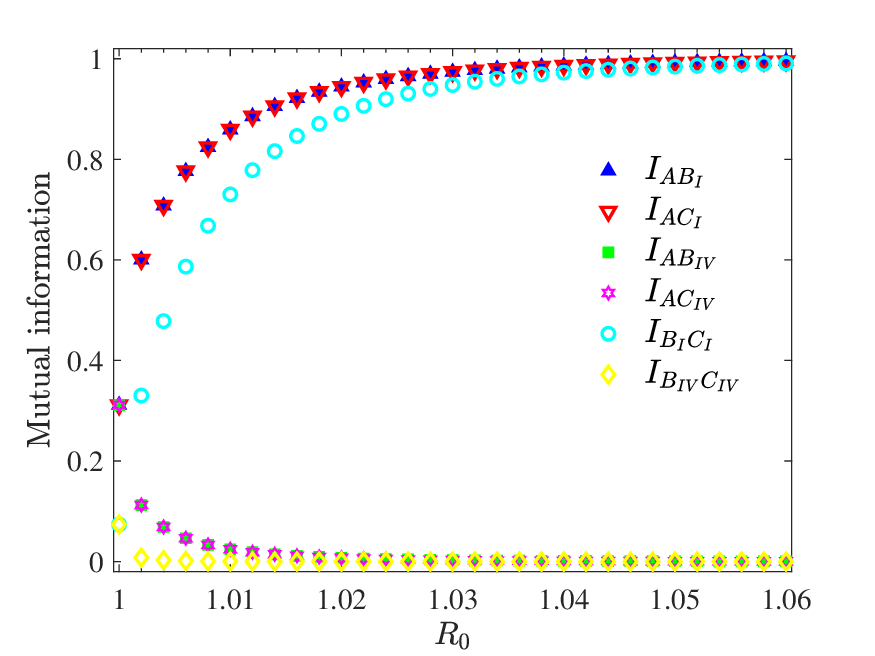}
}
\caption{ Mutual information vs the Hawking temperature $T$ and distance $R_0$ for the initial Werner states in the Schwarzschild black hole with $p$ = 0.5, ${z^2} = \frac{1}{2}$. Graph (a) shows the mutual information vs the Hawking temperature $T$ with $R_0$ = 1.05; Graph (b) presents the mutual information vs the distance $R_0$ with $\Omega = \omega_k / T = 30$.}
\label{fig:6}
\end{figure*}

In order to illustrate the performance of our QMA-EUR, we initially prepare a tripartite state ${\hat\rho _{ABC}}$. In this three-qubit system, particle $A$ remains at the asymptotically flat region, while particles $B$ and $C$ fall towards the Schwarzschild black hole and locate at a fixed distance ${r_0}$ outside the event horizon. Thereby, the system evolves into the state ${\hat\rho _{A{B\rm{_I}}{B\rm{_{IV}}}{C\rm{_I}}{c\rm{_{IV}}}}}$. By tracing out the physically inaccessible part of the composite system's state ${\hat\rho _{A{B\rm{_I}}{B\rm{_{IV}}}{C\rm{_I}}{C\rm{_{IV}}}}}$, one can obtain the physically accessible part of the state ${\hat\rho _{A{B\rm{_I}}{C\rm{_I}}}}$. Here, we employ the two-dimensional Pauli operators ${\hat \sigma _x},{\hat \sigma _y},{\hat \sigma _z}$ as incompatible measurements. Thus, we have  ${q_{MU}} =  - {\log _2}c(\hat R,\hat Q)=1$ with $\hat R$ and $\hat Q$ $\in \left\{ {{{\hat \sigma }_x},{{\hat \sigma }_y},{{\hat \sigma }_z}} \right\}$.

According to eqs. (\ref{Eq.201}) and (\ref{Eq.202}), the following inequalities are given by
\begin{equation}
\left\{ {\begin{array}{*{20}{c}}
{S({{\hat \sigma }_x}|B) + S({{\hat \sigma }_y}|B) \ge 1 + S(A|B)},\\
{S({{\hat \sigma }_x}|B) + S({{\hat \sigma }_z}|C) \ge 1},\\
{S({{\hat \sigma }_y}|B) + S({{\hat \sigma }_z}|C) \ge 1}.
\label{Eq.409}
\end{array}} \right.
\end{equation}
By summing the above inequalities and dividing both sides by $2$, one can obtain
\begin{equation}
U \ge \frac{3}{2} + \frac{1}{2}S(A|B) = \rm Bound_1,
\label{Eq.410}
\end{equation}
where the uncertainty is denoted as $U = S({{\hat \sigma }_x}|B) + S({{\hat \sigma }_y}|B) + S({{\hat \sigma }_z}|C)$.
In addition, we let the subset ${S_1}$ to be ${S_1} = \{ {\hat\sigma _x},{\hat\sigma _y}\}$ and subset $S_2$ to be ${S_2} = \{{\hat\sigma _z}\}$,  then we substitute these into eq. (\ref{Eq.207}) and gain
\begin{equation}
{U \ge \frac{3}{2} + \frac{1}{2}S(A|B) + \max \{ 0,\Delta \} = \rm Bound_2,}
\label{Eq.411}
\end{equation}
with $\Delta  = S(A) - \frac{1}{2}[H({{\hat M}_x}:B) + H({{\hat M}_y}:B)] - H({{\hat M}_z}:C)$.

In order to concreatly represent the entropic uncertainty relations eqs. (\ref{Eq.410}) and (\ref{Eq.411}) in curved spacetime, we first introduce GHZ-type state density matrix
\begin{equation}
{\hat \rho _{ABC}} = \left| {\Psi _{\rm GHZ}} \right\rangle \left\langle {\Psi _{\rm GHZ}} \right|,
\label{Eq.412}
\end{equation}
with $\left| {{\Psi _{\rm GHZ}}} \right\rangle  = z\left| {000} \right\rangle  + \sqrt {1 - {z^2}} \left| {111} \right\rangle $. With $z$ is the state parmeter. Before evolution, the three-qubit GHZ-type state could be expressed as $\left| \Psi_{\text{GHZ}} \right\rangle = z \left| 000 \right\rangle_{ABC} + \sqrt{1-z^2} \left| 111\right\rangle_{ABC}$. When particles $B$ and $C$ hover near the event horizon, we can see by substituting eqs. (\ref{Eq.314}) and (\ref{Eq.315}) that the system evolves into
\begin{equation}
\begin{split}
\left| {{\Psi '}_{\text{GHZ}}} \right\rangle &= z \left( {1 + a} \right)^{-1} \left| {00000} \right\rangle_{A{B_{\text{I}} B_{\text{IV}}}{C_{\text{I}} C_{\text{IV}}}} \\
&\quad + z\left(1 + b \right)^{-1}\left| {0 1 1 1 1} \right\rangle_{A{B_{\text{I}} B_{\text{IV}}}{C_{\text{I}} C_{\text{IV}}}} \\
&\quad + z\left(2 + a + b \right)^{-\frac{1}{2}}\left| {0 0 0 1 1} \right\rangle_{A{B_{\text{I}} B_{\text{IV}}}{C_{\text{I}} C_{\text{IV}}}} \\
&\quad +  z\left(2 + a + b \right)^{-\frac{1}{2}}\left| {0 1 1 0 0} \right\rangle_{A{B_{\text{I}} B_{\text{IV}}}{C_{\text{I}} C_{\text{IV}}}} \\
&\quad + \sqrt {1 - z^2} \left| {1 1 0 1 0} \right\rangle_{A{B_{\text{I}} B_{\text{IV}}}{C_{\text{I}} C_{\text{IV}}}},
\label{Eq.413}
\end{split}
\end{equation}

with $a$ = ${{e^{ - \Omega \sqrt {1 - \frac{1}{R_0}} }}}$, $b$ = ${{e^{ \Omega \sqrt {1 - \frac{1}{R_0}} }}}$ and the evolved density matrix can be expressed as
\begin{equation}
{\hat \rho _{A{B\rm{_I}}{B\rm{_{IV}}}{C\rm{_I}}{C\rm{_{IV}}}}^{\rm GHZ} = {\left| {\Psi' _{\rm GHZ}} \right\rangle}{\left\langle {\Psi' _{\rm GHZ}} \right|}}.
\label{Eq.414}
\end{equation}
In order to obtain the density matrix in the physically accessible region, it's necessary to take the partial trace of the evolved density matrix
\begin{equation}
\hat \rho _{A{B\rm{_I}}{C\rm{_I}}}^{\rm GHZ} = {\rm Tr}_{{B\rm{_{IV}}}{C\rm{_{IV}}}}(\hat \rho _{A{B\rm{_I}}{B\rm{_{IV}}}{C\rm{_I}}{C\rm{_{IV}}}}^{\rm GHZ}).
\label{Eq.415}
\end{equation}
To observe the entropy uncertainty in curved spacetime, the derived uncertainty relations can be obtained by virtue of eqs. (\ref{Eq.410}) and (\ref{Eq.411}). Besides, figure \ref{fig:3} depicts the entropic uncertainty and coherence of the evolved state in the physically accessible region under a Schwarzschild black hole as a function of distance $R_0$ and Hawking Temperature $T$. Following figure \ref{fig:3}, several interesting conclusions are drawn as: (i) The magnitude of the uncertainty is  greater or equal to its bounds all the time, directly indicating the inequalities (\ref{Eq.410}) and (\ref{Eq.411}) we derived are held;  (ii) Bound2 $\ge$ Bound1 is maintained all the time, which implies our bound is tighter than one proposed previously \cite{{JMR}}. That is, our derived bound outperforms the previous. (iii) There exhibits the anti-correlation relationship between the entropic uncertainty and coherence. To be explicit,  both of the uncertainty and its lower bound  decrease  and the $l_1$-norm coherence increases as the distance $R_0$ increases, shown in figures \ref{fig:3}(a) and \ref{fig:3}(c). This phenomenon can be attributed to the fact that the information loss induced by Hawking radiation near the event horizon diminishes with increasing radial distance from the black hole's central singularity, consequently enhancing the quantum coherence of the system.  While, the uncertainty grows and the coherence reduces, as the growing Hawking Temperature $T$ in figures \ref{fig:3}(b) and \ref{fig:3}(d). This is because that  the evaporation of information will be accelerated   as the Hawking temperature rises, which leads to weakening quantumness of the concerned system.
In order to further probe the proposed uncertainty relation within the current framework, we resort to a class of mixed states, i.e., three-qubit Werner states with form of
\begin{equation}
\hat\rho _{ABC}^{{\rm Werner}} = p\left| {\Psi _{\rm GHZ}} \right\rangle \left\langle {\Psi _{\rm GHZ}} \right| + \frac{{(1 - p)}}{8} I {_{8 \times 8}},
\label{Eq.416}
\end{equation}
as the initial states, where $p \in [0,1]$ is denoted as the purity of the system,  and $I {_{8 \times 8}}$ represents an ${8 \times 8}$ density matrix. As a consequence, the evolution of the system can be expressed as
\begin{equation}
\begin{split}
\hat \rho_{A{B_{\mathrm{I}}}{B_{\mathrm{IV}}}{C_{\mathrm{I}}}{C_{\mathrm{IV}}}}^{\rm Werner} 
  &= p{\left| {\Psi' _{\rm GHZ}} \right\rangle  }\langle {\Psi' _{\rm GHZ}}{| }  + \frac{{(1 - p)}}{8}{I_A} \otimes {I_{{B_{\mathrm{I}}}{B_{\mathrm{IV}}}}} \otimes {I_{{C_{\mathrm{I}}}{C_{\mathrm{IV}}}}}\\
&= p{\left| {\Psi' _{\rm GHZ}} \right\rangle  }\langle {\Psi' _{\rm GHZ}}{| }  + \frac{{(1 - p)}}{8}{I_A}\\
&\quad \otimes [({\cal P}\left| {{0_{{B_{\mathrm{I}}}}}} \right\rangle \left| {{0_{{B_{\mathrm{IV}}}}}} \right\rangle  + {\cal Q}\left| {{1_{{B_{\mathrm{I}}}}}} \right\rangle \left| {{1_{{B_{\mathrm{IV}}}}}} \right\rangle )\\
&\quad\quad \otimes ({\cal P}\left\langle {{0_{{B_{\mathrm{I}}}}}} \right|\left\langle {{0_{{B_{\mathrm{IV}}}}}} \right| + {\cal Q}\left\langle {{1_{{B_{\mathrm{I}}}}}} \right|\left\langle {{1_{{B_{\mathrm{IV}}}}}} \right|)\\
&\quad\quad + (\left| {{1_{{B_{\mathrm{I}}}}}} \right\rangle \left| {{0_{{B_{\mathrm{IV}}}}}} \right\rangle )\otimes (\left\langle {{1_{{B_{\mathrm{I}}}}}} \right|\left\langle {{0_{{B_{\mathrm{IV}}}}}} \right|)]\\
&\quad \otimes [({\cal P}\left| {{0_{{C_{\mathrm{I}}}}}} \right\rangle \left| {{0_{{C_{\mathrm{IV}}}}}} \right\rangle  + {\cal Q}\left| {{1_{{C_{\mathrm{I}}}}}} \right\rangle \left| {{1_{{C_{\mathrm{IV}}}}}} \right\rangle )\\
&\quad\quad \otimes ({\cal P}\left\langle {{0_{{C_{\mathrm{I}}}}}} \right|\left\langle {{0_{{C_{\mathrm{IV}}}}}} \right| + {\cal Q}\left\langle {{1_{{C_{\mathrm{I}}}}}} \right|\left\langle {{1_{{C_{\mathrm{IV}}}}}} \right|)\\
&\quad\quad  + (\left| {{1_{{C_{\mathrm{I}}}}}} \right\rangle \left| {{0_{{C_{\mathrm{IV}}}}}} \right\rangle ) \otimes (\left\langle {{1_{{C_{\mathrm{I}}}}}} \right|\left\langle {{0_{{C_{\mathrm{IV}}}}}} \right|)],
\label{Eq.417}
\end{split}
\end{equation}
with intermediate variable ${\cal P} = {\left[ {1 + {e^{\left( { - \Omega \sqrt {1 - {1 \mathord{\left/ {\vphantom {1 {{R_0}}}} \right.\kern-\nulldelimiterspace} {{R_0}}}} } \right)}}} \right]^{ - \frac{1}{2}}}$ and ${\cal Q} = {\left[ {1 + {e^{\left( {\Omega \sqrt {1 - {1 \mathord{\left/ {\vphantom {1 {{R_0}}}} \right. \kern-\nulldelimiterspace} {{R_0}}}} } \right)}}} \right]^{ - \frac{1}{2}}}$. 
By taking the partial trace of evolved system $\hat \rho _{A{B_I}{B_{IV}}{C_I}{C_{IV}}}^{\rm Werner}$,  the reduced state of the physically accessible can be given by
\begin{equation}
\hat \rho _{A{B\rm{_I}}{C\rm{_I}}}^{\rm Werner} = {\rm Tr}_{{B\rm{_{IV}}}{C\rm{_{IV}}}}(\hat \rho _{A{B\rm{_I}}{B\rm{_{IV}}}{C\rm{_I}}{C\rm{_{IV}}}}^{\rm Werner}).
\label{Eq.418}
\end{equation}
For clarity, the analytical expressions  of the uncertainty and its bounds for the GHZ state and the Werner state are offered in Appendix \ref{appendixC} and Appendix \ref{appendixD}, respectively. Figure \ref{fig:4} has plotted the uncertainty and two bounds as functions of Hawking temperature $T$ and the distance $R_0$ when the system  is initially prepared in the three-qubit Werner state. 
Similar to the former case, one can attain the same conclusions, including the availability and optimization of our derived QMA-EUR (Eq.  (\ref{Eq.207})), and the anti-correlation relationship between the uncertainty and the system's coherence. Noteworthily, there are two distinct features  for the cases in the two different initial states (three-qubit GHZ-type and Werner states) as: (i) The uncertainty in the current consideration is closer to the uncertainty bound. This manifests that the lower bound can better reflect the nature of uncertainty in the latter, which is related to the von Neumann entropy and Holevo quantity, shown as the right item of Eq. (\ref{Eq.411}). (ii) Although the uncertainty and coherence are eventually stabilized in figures \ref{fig:3} and \ref{fig:4}, however, those in figure \ref{fig:4} stabilize faster than the case in the former case. This can be explained as that the purity of the Werner state is going down more rapidly compared with that of the pure GHZ-type state in the black hole. Furthermore, 
Fig. \ref{fig:5} has drawn the bound gain ($\Delta={\rm Bound}2-{\rm Bound}1$) as functions of $R_0$, $T$ and $z$ in tripartite GHZ-type and Werner states. It demonstrates that ${\Delta \ge 0}$ remains always. This positive gain underscores the robustness and general applicability of our tighter entropic uncertainty relation in the curved spacetime.

To further clarify the decoherence induced by black hole and the anti-correlation between the uncertainty and $l_1$-norm coherence, we resort to the mutual information $I$  related to the Hawking temperature $T$ and distance $R_0$, which has been indicated in Fig. \ref{fig:6}.   Following the figure, the mutual information $I_{AB_I}$, $I_{AC_I}$ and $I_{B_IC_I}$ in  physically accessible region decreases as the increasing Hawking temperature and decreasing distance while  the  mutual information $I_{AB_{IV}}$, $I_{AC_{IV}}$ and $I_{B_{IV}C_{IV}}$ in physically inaccessible region exhibits the opposite trends in the current consideration. This indicates that the black hole causes a redistribution of information, resulting in that its information in the physically accessible region flows into the physically inaccessible region, which leads to the decoherence and raise the system's intrinsic uncertainty.  

\section{Conclusion}
\label{sec5}
In this article, we have derived a new and universal QMA-EUR for $m$ measurements within a $n+1$-particle system. Specifically, we observe the derived EUR and quantumness in the background of the Schwarzschild spacetime, where the tripartite GHZ-type and Werner states are served as the initial state respectively. It has been found that our proposed EUR are held always, and its lower bound outperforms the previous ones.  Besides, the growing Hawking temperature $T$ induces the inflation of the uncertainty and the reduction of the $l_1$-norm coherence in the physically accessible region.  While, the effect  of the distance $R_0$ between the particle and the center of the black hole is completely opposite compared with that raised by the Hawking temperature. With these in mind, it reveals that the uncertainty is anti-correlated with the $l_1$-norm coherence. 
Furthermore, the $l_1$-norm coherence and concurrence of the $N$-partite GHZ-type states have been examined in details, in order to uncover the quantumness for the whole evolved system with $n$ observers hovering near the event horizon and $N-n$ in the  flat region. Remarkably, it is found that the coherence and entanglement  of the system possess the identical expressions in the current scenario. Therefore, our investigations shed light on the generalized entropic uncertainty relation and quantumness in curved spacetimes.

\acknowledgments

This work was supported by the National Science
Foundation of China under (Grants No. 12475009 and
No. 12075001), Anhui Provincial Key Research and
Development Plan (Grant No. 2022b13020004), Anhui
Province Science and Technology Innovation Project
(Grant No. 202423r06050004), Anhui Provincial
University Scientific Research Major Project (Grant
No. 2024AH040008), and Anhui Province Natural Science Foundation (Grant No. 202508140141).

\appendix
\begin{widetext}
\hypertarget{A}{\section{Derivation of the generalized EUR}}
\label{appendixA}

This appendix provides a detailed derivation of the coefficients appearing in the generalized entropic uncertainty relation (EUR) presented in Eq. (\ref{Eq.207}) of the main text, thereby enhancing the transparency of the proof.

\subsection*{Derivation Process}

The derivation of the generalized EUR (Eq. (\ref{Eq.207})) proceeds through systematic summation of pairwise uncertainty relations. For clarity, we outline the key steps:

1. For each measurement $\hat{M}_i$ in subset $S_t$, we consider all pairwise uncertainty relations with other measurements. These include the relations with measurements in the same subset $S_t$ (using Eq. (\ref{Eq.201})) and the relations with measurements in different subsets $S_s$ $(s\neq t)$ (using Eq. (\ref{Eq.205}))

2. Summation for each measurement: for the quantum memory $B_1$ and fixed measurement $\hat{M}_1 \in S_1$, we sum all inequalities involving $\hat{M}_1$ and other measurements (Eq. (\ref{Eq.208}) and Eq. (\ref{Eq.209})), then normalize by $(m-1)$ to yield an inequality for $B_1$ and $\hat{M}_1$ (Eq. (\ref{Eq.210})).

3. Summation for each subset: we repeat this process for all measurements in $S_1$, obtaining inequalities for $\hat{M}_2, \dots, \hat{M}_{m_1}$ (Eqs. (\ref{Eq.211}-\ref{Eq.212}). Summing these $m_1$ inequalities yields the consolidated inequality for subset $S_1$ (Eq. (\ref{Eq.213})).

4. Extension to all subsets: Following the same methodology, we derive similar inequalities for subsets $S_2, \dots, S_{n-1}$ (Eq. (\ref{Eq.214}) and its generalizations) and for the final subset $S_n$ (Eq. (\ref{Eq.215})). After summing the inequalities for all $n$ subsets and  normalizing, we attain the generalized EUR (i.e., Eq. (\ref{Eq.207})).

\subsection*{Coefficient Analysis}
The coefficients in Eq. (\ref{Eq.207})  stem from the combinatorial structure associated with the pairwise summations. We present the complete coefficients for all terms of the generalized EUR in Table {\ref{tab1}, where LHS represents the left-hand side of the inequality, and RHS represents the right-hand side of the  inequality.}
\renewcommand{\arraystretch}{1.5}
\begin{table}[h!]
\centering
\caption{Complete coefficients for the generalized EUR (Eq.~(\ref{Eq.207}))}
\begin{tabularx}{\linewidth}{l>{\centering}Xc}
\hline
\textbf{Term types} & \textbf{Terms} & \textbf{Coefficients} \\
\hline
Uncertainty (LHS)  & $S(\hat{M}_i|B_t)$ for $\hat{M}_i \in S_t$ and $t = 1,2,...,n$ & $m -1$ \\
\hline
Logarithmic terms in RHS & $-\log_2 c_{ij}$ for $i = 1,2,...,m - 1 $ and $j = i + 1,...,m$ & 1 \\
\hline
& $S(A|B_1)$ & $\displaystyle \frac{m_1(m_1-1)}{2}$ \\
Conditional entropies in RHS & $S(A|B_2)$ & $\displaystyle \frac{m_2(m_2-1)}{2}$ \\
 & $\vdots$ & $\vdots$ \\
& $S(A|B_n)$ & $\displaystyle \frac{m_n(m_n-1)}{2}$ \\
\hline
Subsystem entropy in RHS & $S(A)$ & $\sum\limits_{i = 1}^{n - 1} m_i \left( \sum\limits_{j = i + 1}^n m_j \right)$ \\
\hline
Holevo quantities in RHS& $H(\hat{M}_i:B_t)$ for $\hat{M}_i \in S_t$ and $t = 1,2,...,n$ & $\sum\limits_{t = 1}^n \sum\limits_{{M_i} \in {S_t}} \left( \sum\limits_{j = 1,j \ne t}^n m_j \right)$  \\
\hline
\end{tabularx}
\label{tab1}
\end{table}
\renewcommand{\arraystretch}{1.0}

\hypertarget{A}{\section{Sub-matrixes of evolved system}}
\label{appendixB}
The evolved system {$\hat \rho _{N - n,p,q}$} can be expressed as:
\begin{equation}
{\hat \rho _{N - n,p,q}} = \left( {\begin{array}{*{20}{c}}
A&C\\
{{C^T}}&B
\end{array}} \right).
\label{Eq.a1}
\end{equation}
From  the matrix in Eq. (\ref{Eq.401}), it can be seen that there is the only non-zero element ${b_{{2^q}}} = 1 - {z^2}$, and the sub-matrix $B$ of Eq. (\ref{Eq.406}) is given by
\begin{equation}
B = 
\left(
  \begin{array}{ccccccc}
    0 & & & & & & \\
     & \ddots & & & & & \\
     & & 0 & & & & \\
     & & & 1 - z^2 & & & \\
     & & & & 0 & & \\
     & & & & & \ddots & \\
     & & & & & & 0
  \end{array}
\right)
\raisebox{2.5em}{
  $\left.\vphantom{
    \begin{array}{c}
      0 \\
      \ddots \\
      0 \\
      1 - z^2
    \end{array}
  }\right\rbrace$ $\displaystyle 2^q\ {\rm items}$
}
\label{Eq.a2}
\end{equation}
Similarly, only ${c_{{2^q}}} = \frac{{z\sqrt {1 - {z^2}} }}{{\sqrt {{{({e^{ - \Omega \sqrt {{{1 - 1} \mathord{\left/ {\vphantom {{1 - 1} {{R_0}}}} \right. \kern-\nulldelimiterspace} {{R_0}}}} }} + 1)}^p}{{({e^{\Omega \sqrt {{{1 - 1} \mathord{\left/ {\vphantom {{1 - 1} {{R_0}}}} \right. \kern-\nulldelimiterspace} {{R_0}}}} }} + 1)}^q}} }}$ is not zero and the sub-matrix $C$ is written as
\begin{equation}
C = \left( {\begin{array}{*{20}{c}}
{}&{}&{}&{}&{}&{}&0\\
{}&{}&{}&{}&{}& {\mathinner{\mkern2mu\raise1pt\hbox{.}\mkern2mu
 \raise4pt\hbox{.}\mkern2mu\raise7pt\hbox{.}\mkern1mu}} &{}\\
{}&{}&{}&{}&0&{}&{}\\
{}&{}&{}&{\frac{{z\sqrt {1 - {z^2}} }}{{\sqrt {{{({e^{ - \Omega \sqrt {1 - 1/{R_0}} }} + 1)}^p}{{({e^{\Omega \sqrt {1 - 1/{R_0}} }} + 1)}^q}} }}}&{}&{}&{}\\
{}&{}&0&{}&{}&{}&{}\\
{}& {\mathinner{\mkern2mu\raise1pt\hbox{.}\mkern2mu
 \raise4pt\hbox{.}\mkern2mu\raise7pt\hbox{.}\mkern1mu}} &{}&{}&{}&{}&{}\\
0&{}&{}&{}&{}&{}&{}
\end{array}} \right)
\raisebox{2.0em}{
  $\left.\vphantom{
    \begin{array}{c}
      0 \\
      \ddots \\
      0 \\
      1 - z^2
    \end{array}
  }\right\rbrace$ $\displaystyle 2^q\ {\rm items}$
}
\label{Eq.a3}
\end{equation}

\hypertarget{A}{\section{Analytical expressions for the uncertainty and its bounds in Schwarzschild black hole when the initial state of the system is in the state of the tripartite GHZ-type state}}
\label{appendixC}
The uncertainty can be expressed as:
\begin{equation}
\begin{aligned}
U &= 3 \times \log_2\left[1 - z^2 + \frac{z^2}{(E_p + 1)^2} + 4K^2 z^2\right] \times \left[1 - z^2 + \frac{z^2}{(E_p + 1)^2} + 4K^2 z^2\right] \\
&\quad - 4 \times \log_2\left[\frac{1 - z^2}{2} + \frac{z^2}{2(E_p + 1)^2} + 2K^2 z^2\right] \times \left[\frac{1 - z^2}{2} + \frac{z^2}{2(E_p + 1)^2} + 2K^2 z^2\right] \\
&\quad + 2 \times \log_2\left[\frac{z^2}{(E_n + 1)^2} + 4K^2 z^2\right] \times \left[\frac{z^2}{(E_n + 1)^2} + 4K^2 z^2\right] \\
&\quad - 4 \times \log_2\left[\frac{z^2}{2(E_n + 1)^2} + 2K^2 z^2\right] \times \left[\frac{z^2}{2(E_n + 1)^2} + 2K^2 z^2\right] \\
&\quad - \log_2\left[\frac{z^2}{(E_p + 1)^2} + 4K^2 z^2\right] \times \left[\frac{z^2}{(E_p + 1)^2} + 4K^2 z^2\right] \\
&\quad - \log_2\left(1 - z^2\right) \times (1 - z^2)
\end{aligned}
\label{Eq.b1}
\end{equation}
where  ${E_p = e^{\frac{w_k}{T} \sqrt{1 - \frac{1}{r}}}}$ ,  ${E_n = e^{-\frac{w_k}{T} \sqrt{1 - \frac{1}{r}}}}$, and ${K = \frac{1}{2\sqrt{(E_p + 1)(E_n + 1)}}}$ are intermediate variables.

And the analytical expression of the Bound1 is given by
\begin{equation}
\begin{aligned}
\text{Bound1} &= \frac{1}{2} \times \log_2\left[ (1-z^2) + 4\varepsilon z^2 \right] \times \left[ (1-z^2) + 4\varepsilon z^2 \right]  - \frac{1}{2} \times \log_2\left( 4\varepsilon z^2 \right) \times \left( 4\varepsilon z^2 \right) \\
&\quad - \frac{1}{2} \times \log_2\left( 1-z^2 \right) \times (1-z^2) + \frac{3}{2}
\end{aligned}
\label{Eq.b2}
\end{equation}
with ${\varepsilon = \frac{1}{2\left(E_p + 1\right)\left(E_n + 1\right)}}$.
Similarly, one can   obtain the  Bound2 as
\begin{equation}
\begin{aligned}
\text{Bound2} &= \frac{5}{2} \log_2\left(1 - z^2 + 4\varepsilon z^2\right) \times \left(1 - z^2 + 4\varepsilon z^2\right) \\
&\quad - 2 \log_2\left(\frac{1 - z^2}{2} + 2\varepsilon z^2\right) \times \left(\frac{1 - z^2}{2} + 2\varepsilon z^2\right) \\
&\quad + \log_2\left[\frac{z^2}{(E_n + 1)^2} + 2\varepsilon z^2\right] \times \left[\frac{z^2}{(E_n + 1)^2} + 2\varepsilon z^2\right] \\
&\quad - 2 \log_2\left[\frac{z^2}{2(E_n + 1)^2} + \varepsilon z^2\right] \times \left[\frac{z^2}{2(E_n + 1)^2} + \varepsilon z^2\right] \\
&\quad + 2 \log_2\left[\frac{1 - z^2}{2} + \frac{z^2}{2(E_n + 1)^2} + 3\varepsilon z^2\right] \times \left[\frac{1 - z^2}{2} + \frac{z^2}{2(E_n + 1)^2} + 3\varepsilon z^2\right] \\
&\quad - \frac{3}{2} \log_2\left(4\varepsilon z^2\right) \times \left(4\varepsilon z^2\right) - \frac{3}{2} \log_2\left(1 - z^2\right) \times \left(1 - z^2\right) + \frac{3}{2}
\end{aligned}
\label{Eq.b3}
\end{equation}
\hypertarget{A}{\section{Analytical expressions for the uncertainty and its bounds in Schwarzschild black hole when the initial state of the system is in the state of the tripartite Werner state}}\label{appendixD}

In this scenario, the uncertainty can be analytically expressed as 
\begin{align}
U &= \log_2\left[ p(1-z^2) - \frac{(p-1)}{4(E_p + 1)} - \frac{(p-1)}{8(E_p + 1)^2} - \frac{(p-1)}{8(E_n + 1)} - \frac{p}{8} - \frac{(p-1)}{8(E_p + 1)(E_n + 1)} + \frac{1}{8} \right]\nonumber \\ 
&\quad \times \left[ \frac{p}{8} + \frac{(p-1)}{4(E_p + 1)} + \frac{(p-1)}{8(E_p + 1)^2} + \frac{(p-1)}{8(E_n + 1)} + \frac{(p-1)}{8(E_p + 1)(E_n + 1)} - p(1-z^2) - \frac{1}{8} \right]\nonumber \\
&\quad + \log_2\left[ \frac{p z^2}{(E_p + 1)^2} - \frac{(p-1)}{4(E_p + 1)} - \frac{(p-1)}{8(E_p + 1)^2} - \frac{(p-1)}{8(E_n + 1)} - \frac{(p-1)}{8(E_p + 1)(E_n + 1)} - \frac{p}{8} + 4K^2 p z^2 + \frac{1}{8} \right]\nonumber \\
&\quad \times \left[ \frac{p}{8} + \frac{(p-1)}{4(E_p + 1)} + \frac{(p-1)}{8(E_p + 1)^2} + \frac{(p-1)}{8(E_n + 1)} + \frac{(p-1)}{8(E_p + 1)(E_n + 1)} - \frac{p z^2}{(E_p + 1)^2} - 4K^2 p z^2 - \frac{1}{8} \right]\nonumber \end{align}
\begin{align}
&\quad + \log_2\left[ \frac{p z^2}{(E_n + 1)^2} - \frac{(p-1)}{8(E_n + 1)^2} - \frac{(p-1)}{8(E_n + 1)} - \frac{(p-1)}{8(E_p + 1)(E_n + 1)} + 4K^2 p z^2 \right] \nonumber \\
&\quad \times \left[ \frac{(p-1)}{8(E_n + 1)} + \frac{(p-1)}{8(E_n + 1)^2} + \frac{(p-1)}{8(E_p + 1)(E_n + 1)} - \frac{p z^2}{(E_n + 1)^2} - 4K^2 p z^2 \right] \nonumber \\
&\quad - 3 \log_2\left[ \frac{p z^2}{(E_n + 1)^2} - \frac{(p-1)}{4(E_n + 1)} - \frac{(p-1)}{4(E_n + 1)^2} - \frac{(p-1)}{4(E_p + 1)(E_n + 1)} + 4K^2 p z^2 \right] \nonumber \\
&\quad \times \left[ \frac{(p-1)}{4(E_n + 1)} + \frac{(p-1)}{4(E_n + 1)^2} + \frac{(p-1)}{4(E_p + 1)(E_n + 1)} - \frac{p z^2}{(E_n + 1)^2} - 4K^2 p z^2 \right] \nonumber \\
&\quad + 4 \log_2\left[ \frac{p z^2}{2(E_n + 1)^2} - \frac{(p-1)}{8(E_n + 1)} - \frac{(p-1)}{8(E_n + 1)^2} - \frac{(p-1)}{8(E_p + 1)(E_n + 1)} + 2K^2 p z^2 \right] \nonumber \\
&\quad \times \left[ \frac{(p-1)}{8(E_n + 1)} + \frac{(p-1)}{8(E_n + 1)^2} + \frac{(p-1)}{8(E_p + 1)(E_n + 1)} - \frac{p z^2}{2(E_n + 1)^2} - 2K^2 p z^2 \right] \nonumber \\
&\quad - 3 \log_2\left[ p(1-z^2) - \frac{(p-1)}{2(E_p + 1)} - \frac{(p-1)}{4(E_p + 1)^2} - \frac{(p-1)}{4(E_n + 1)} - \frac{p}{4} - \frac{(p-1)}{4(E_p + 1)(E_n + 1)} + \frac{p z^2}{(E_p + 1)^2} + 4K^2 p z^2 + \frac{1}{4} \right] \nonumber \\
&\quad \times \left[ \frac{p}{4} + \frac{(p-1)}{2(E_p + 1)} + \frac{(p-1)}{4(E_p + 1)^2} + \frac{(p-1)}{4(E_n + 1)} + \frac{(p-1)}{4(E_p + 1)(E_n + 1)} - p(1-z^2) - \frac{p z^2}{(E_p + 1)^2} - 4K^2 p z^2 - \frac{1}{4} \right] \nonumber \\
&\quad + 4 \log_2\left[ \frac{p(1-z^2)}{2} - \frac{(p-1)}{4(E_p + 1)} - \frac{(p-1)}{8(E_p + 1)^2} - \frac{(p-1)}{8(E_n + 1)} - \frac{p}{8} - \frac{(p-1)}{8(E_p + 1)(E_n + 1)} + \frac{p z^2}{2(E_p + 1)^2} + 2K^2 p z^2 + \frac{1}{8} \right] \nonumber \\
&\quad \times \left[ \frac{p}{8} + \frac{(p-1)}{4(E_p + 1)} + \frac{(p-1)}{8(E_p + 1)^2} + \frac{(p-1)}{8(E_n + 1)} + \frac{(p-1)}{8(E_p + 1)(E_n + 1)} - \frac{p(1-z^2)}{2} - \frac{p z^2}{2(E_p + 1)^2} - 2K^2 p z^2 - \frac{1}{8} \right] \nonumber \\
&\quad + \log_2\left[ - \frac{(p-1)}{8(E_n + 1)} - \frac{(p-1)}{8(E_n + 1)^2} - \frac{(p-1)}{8(E_p + 1)(E_n + 1)} \right] \nonumber \\
&\quad \times \left[ \frac{(p-1)}{8(E_n + 1)} + \frac{(p-1)}{8(E_n + 1)^2} + \frac{(p-1)}{8(E_p + 1)(E_n + 1)} \right]
\end{align}
with ${E_p = e^{\frac{w_k}{T} \sqrt{1 - \frac{1}{r}}}}$,
${E_n = e^{-\frac{w_k}{T} \sqrt{1 - \frac{1}{r}}}}$,
and ${K = \frac{1}{2\sqrt{(E_p + 1)(E_n + 1)}}}$.

Moreover, the Bound1 and Bound2 in the current scenario can be calculated as
\begin{align}
\text{Bound1} &= \frac{1}{2} \log_2(A_1) \times B_1 + \frac{1}{2} \log_2(A_3) \times B_3 + \frac{1}{2} \log_2(A_4) \times B_4 \nonumber \\
&\quad - \frac{1}{2} \log_2(A_5) \times B_5 - \frac{1}{2} \log_2(A_7) \times B_7 + \frac{1}{2} \log_2(A_9) \times B_9 + \frac{3}{2}, \\
\nonumber \\
\text{Bound2} &= \frac{3}{2} \log_2(A_1) \times B_1 - 2 \log_2(A_2) \times B_2 + \frac{3}{2} \log_2(A_3) \times B_3 + \frac{3}{2} \log_2(A_4) \times B_4 \nonumber \\
&\quad - \frac{5}{2} \log_2(A_5) \times B_5 + 2 \log_2(A_6) \times B_6 - \frac{5}{2} \log_2(A_7) \times B_7 + 2 \log_2(A_8) \times B_8 \nonumber \\
&\quad + \frac{3}{2} \log_2(A_9) \times B_9 + \frac{3}{2}
\end{align}
respectively, with  
\begin{align*}
q &= 1-p, \\
A_1 &= p(1 - z^2) - \frac{2q}{E_p + 1} - \frac{q}{(E_p + 1)^2} - \frac{q}{E_n + 1} - \frac{p}{8} - \frac{q}{(E_p + 1)(E_n + 1)} + \frac{1}{8}, \\
B_1 &= \frac{p}{8} + \frac{2q}{E_p + 1} + \frac{q}{(E_p + 1)^2} + \frac{q}{E_n + 1} - p(1 - z^2) + \frac{q}{(E_p + 1)(E_n + 1)} - \frac{1}{8}, \end{align*}
\begin{align*}
A_2 &= \frac{p(1 - z^2)}{2} - \frac{2q}{E_p + 1} - \frac{q}{(E_p + 1)^2} - \frac{2q}{E_n + 1} - \frac{q}{(E_n + 1)^2} - \frac{p}{8}\\
&\quad - \frac{2q}{(E_p + 1)(E_n + 1)} + \frac{p z^2}{2(E_p + 1)^2} + \frac{p z^2}{2(E_n + 1)^2} + p z^2 + \frac{1}{8}, \\
B_2 &= \frac{p}{8} + \frac{2q}{E_p + 1} + \frac{q}{(E_p + 1)^2} + \frac{2q}{E_n + 1} + \frac{q}{(E_n + 1)^2} - \frac{p(1 - z^2)}{2},\\
&\quad + \frac{2q}{(E_p + 1)(E_n + 1)} - \frac{p z^2}{2(E_p + 1)^2} - \frac{p z^2}{2(E_n + 1)^2} - p z^2 - \frac{1}{8}, \\
A_3 &= \frac{p z^2}{(E_p + 1)^2} - \frac{2q}{E_p + 1} - \frac{q}{(E_p + 1)^2} - \frac{q}{E_n + 1} - \frac{q}{(E_p + 1)(E_n + 1)} - \frac{p}{8} + p z^2 + \frac{1}{8}, \\
B_3 &= \frac{p}{8} + \frac{2q}{E_p + 1} + \frac{q}{(E_p + 1)^2} + \frac{q}{E_n + 1} + \frac{q}{(E_p + 1)(E_n + 1)} - \frac{p z^2}{(E_p + 1)^2} - p z^2 - \frac{1}{8},\\
A_4 &= \frac{p z^2}{(E_n + 1)^2} - \frac{q}{(E_n + 1)^2} - \frac{q}{(E_p + 1)(E_n + 1)} - \frac{q}{E_n + 1} + p z^2,  \\
B_4 &= \frac{q}{E_n + 1} + \frac{q}{(E_n + 1)^2} + \frac{q}{(E_p + 1)(E_n + 1)} - \frac{p z^2}{(E_n + 1)^2} - p z^2, \\
A_5 &= \frac{p z^2}{(E_n + 1)^2} - \frac{2q}{(E_n + 1)^2} - \frac{2q}{(E_p + 1)(E_n + 1)} - \frac{2q}{E_n + 1} + p z^2, \\
B_5 &= \frac{2q}{E_n + 1} + \frac{2q}{(E_n + 1)^2} + \frac{2q}{(E_p + 1)(E_n + 1)} - \frac{p z^2}{(E_n + 1)^2} - p z^2, \\
A_6 &= \frac{p z^2}{2(E_n + 1)^2} - \frac{q}{(E_n + 1)^2} - \frac{q}{(E_p + 1)(E_n + 1)} - \frac{q}{E_n + 1} + \frac{p z^2}{2}, \\
B_6 &= \frac{q}{E_n + 1} + \frac{q}{(E_n + 1)^2} + \frac{q}{(E_p + 1)(E_n + 1)} - \frac{p z^2}{2(E_n + 1)^2} - \frac{p z^2}{2}, \\
A_7 &= p(1 - z^2) - \frac{4q}{E_p + 1} - \frac{2q}{(E_p + 1)^2} - \frac{2q}{E_n + 1} - \frac{p}{4} - \frac{2q}{(E_p + 1)(E_n + 1)} + \frac{p z^2}{(E_p + 1)^2} + p z^2 + \frac{1}{4}, \\
B_7 &= \frac{p}{4} + \frac{4q}{E_p + 1} + \frac{2q}{(E_p + 1)^2} + \frac{2q}{E_n + 1} - p(1 - z^2) + \frac{2q}{(E_p + 1)(E_n + 1)} - \frac{p z^2}{(E_p + 1)^2} - p z^2 - \frac{1}{4}, \end{align*}
\begin{align*}
A_8 &= \frac{p(1 - z^2)}{2} - \frac{2q}{E_p + 1} - \frac{q}{(E_p + 1)^2} - \frac{q}{E_n + 1} - \frac{p}{8} - \frac{q}{(E_p + 1)(E_n + 1)} + \frac{p z^2}{2(E_p + 1)^2} + \frac{p z^2}{2} + \frac{1}{8}, \\
B_8 &= \frac{p}{8} + \frac{2q}{E_p + 1} + \frac{q}{(E_p + 1)^2} + \frac{q}{E_n + 1} - \frac{p(1 - z^2)}{2} + \frac{q}{(E_p + 1)(E_n + 1)} - \frac{p z^2}{2(E_p + 1)^2} - \frac{p z^2}{2} - \frac{1}{8}, \\
A_9 &= - \frac{q}{E_n + 1} - \frac{q}{(E_n + 1)^2} - \frac{q}{(E_p + 1)(E_n + 1)}, \\
B_9 &= \frac{q}{E_n + 1} + \frac{q}{(E_n + 1)^2} + \frac{q}{(E_p + 1)(E_n + 1)}.
\end{align*}

\end{widetext}

\end{document}